\documentclass[aps,prx,twocolumn,footinbib,nobalancelastpage]{revtex4-1}

\usepackage{graphicx}
\usepackage{indentfirst}
\usepackage{physics}
\usepackage{braket}
\usepackage{float}
\usepackage{amsmath}
\usepackage{theorem}
\usepackage{upgreek}
\usepackage{verbatim}
\usepackage{epstopdf}
\usepackage{CJK}
\usepackage{esint}
\usepackage{color}
\usepackage[T1]{fontenc}
\usepackage{subfigure}
\usepackage{amsfonts}
\usepackage{footmisc}
\usepackage{scrextend}
\usepackage{multirow}
\usepackage[english]{babel}
\usepackage{url}
\usepackage{bm}
\usepackage{hyperref}
\definecolor{darkblue}{rgb}{0,0,0.5}
\hypersetup{
colorlinks=true,
linkcolor=black,
filecolor=blue,
citecolor=darkblue,  
urlcolor=black,
}

\newtheorem{theorem}{Theorem}

\newtheorem{lemma}[theorem]{Lemma}

\newcommand{\calE}{{\cal E}}

\newcommand{\calI}{{\cal I}}
\newcommand{\calL}{{\cal L}}

\newcommand{\calH}{{\cal H}}

\newcommand{\1}{^{(1)}}

\def\be{\begin{equation}}
\def\ee{\end{equation}}
\def\ba{\begin{eqnarray}}
\def\ea{\end{eqnarray}}
\usepackage{bm}

\newcommand{\QZ}[1]{{{\textcolor{black}{#1}}}}
\newcommand{\SP}[1]{{{\textcolor{black}{#1}}}}
\newcommand{\SPnew}[1]{{{\textcolor{black}{#1}}}}
\newcommand{\QZnew}[1]{{{\textcolor{black}{#1}}}}

\begin{document}

\title{Entanglement-enhanced testing of multiple quantum hypotheses}

\author{Quntao Zhuang$^{1,2}$}
\email{zhuangquntao@email.arizona.edu}
\author{Stefano Pirandola$^{3,4}$}
\email{stefano.pirandola@york.ac.uk}
\affiliation{
$^1$Department of Electrical and Computer Engineering, University of Arizona, Tucson, AZ 85721, USA
\\
$^2$James C. Wyant College of Optical Sciences, University of Arizona, Tucson, AZ 85721, USA
\\
$^3$Department of Computer Science, University of York, York YO10 5GH, UK
\\
$^4$Research Laboratory of Electronics, Massachusetts Institute of Technology, Cambridge MA 02139, USA
}
\date{\today}

\begin{abstract}

\SP{Quantum hypothesis testing has been greatly advanced for the binary discrimination of two states, or two channels. In this setting, we already know that quantum entanglement can be used to enhance the discrimination of two bosonic channels. Here, we remove the restriction of binary hypotheses and show that entangled photons can remarkably boost the discrimination of multiple bosonic channels. More precisely, we formulate a general problem of channel-position finding where the goal is to determine the position of a target channel among many background channels. We prove that, using entangled photons at the input and a generalized form of conditional nulling receiver at the output, we may outperform any classical strategy. Our results can be applied to enhance a range of technological tasks, including the optical readout of sparse classical data, the spectroscopic analysis of a frequency spectrum, and the determination of the direction of a target at fixed range.}
\end{abstract}

\maketitle

\section{Introduction}
Quantum sensing~\cite{ReviewSensing} exploits quantum resources and measurements to improve the performance of parameter estimation and hypothesis testing, with respect to the best possible classical strategies. One of the fundamental settings of quantum hypothesis testing~\cite{Helstrom_1976,hirota,Anthony_1998,Chefles_2000} is quantum channel discrimination~\cite{KitaevDiamond,Acin_2001,sacchi2005entanglement,wang2006unambiguous,hayashi}, where the aim is to discriminate between different physical processes, modeled as quantum channels, arbitrarily chosen from some known ensemble. Finding the best strategy for quantum channel discrimination is a non-trivial double optimization problem which involves the optimization of both input states and output measurements. 
Furthermore, the optimization is generally performed assuming a certain number of probings and it becomes an energy-constrained problem in the discrimination of bosonic channels, where the available input states have a finite mean number of photons~\cite{weedbrook2012gaussian}.

For the discrimination of bosonic channels, the so-called `classical strategies' are based on preparing the input signal modes in (mixtures of) coherent states and then measuring the channel outputs by means of suitable receivers, e.g., a homodyne detector. By fixing the input energy to a suitably low number of mean photons per probing, the classical strategies are often beaten by truly-quantum sources such as two-mode squeezed vacuum states, where each signal mode (probing the channel) is entangled with a corresponding idler mode directly sent to the output measurement. This quantum advantage was specifically proven for the readout of data from an optical memory, known as quantum reading~\cite{pirandola2011quantum}, and the yes/no detection of a remote target, known as quantum illumination~\cite{tan2008quantum,lloyd2008enhanced,ShabirPRL,nair2020}.  

While quantum advantage with entangled-assisted protocols has been proven in problems of binary quantum channel discrimination with bosonic channels, the potential advantage of quantum entanglement over the best classical strategies still needs to be explored and fully quantified in the more general setting of discrimination between multiple quantum channels. As a matter of fact, this problem is very relevant because real physical applications often involves multiple hypotheses, and their treatment lead to non-trivial mathematical complications. In fact, naively decomposing a multi-hypothesis quantum channel discrimination into multiple rounds of binary cases does not necessarily preserve the quantum advantages from the binary case.

\QZ{In this work, we formulate a basic problem of multiple channel discrimination that we call
``channel-position finding''. Here the goal is to determine the position
of a target channel among many copies of a background channel. We prove that, using entangled photons at
the input and a generalized form of conditional nulling receiver at the output, we may outperform
any classical strategy in finding the position of the target channel, with a clear advantage in terms
of mean error probability and its error exponent. In particular, our receiver design only relies on state-of-the-art technology in quantum optics, i.e., direct photo-detection (not requiring number-resolution), two-mode squeezing (which can be realized by standard optical parametric amplifiers) and feed-forward control (which has been demonstrated~\cite{chen2012optical}). Our results can be applied to various applications, including position-based quantum reading, spectroscopy and target finding.}

\begin{figure}[t!]
\vspace{-0cm}
\centering\includegraphics[width=0.45\textwidth]{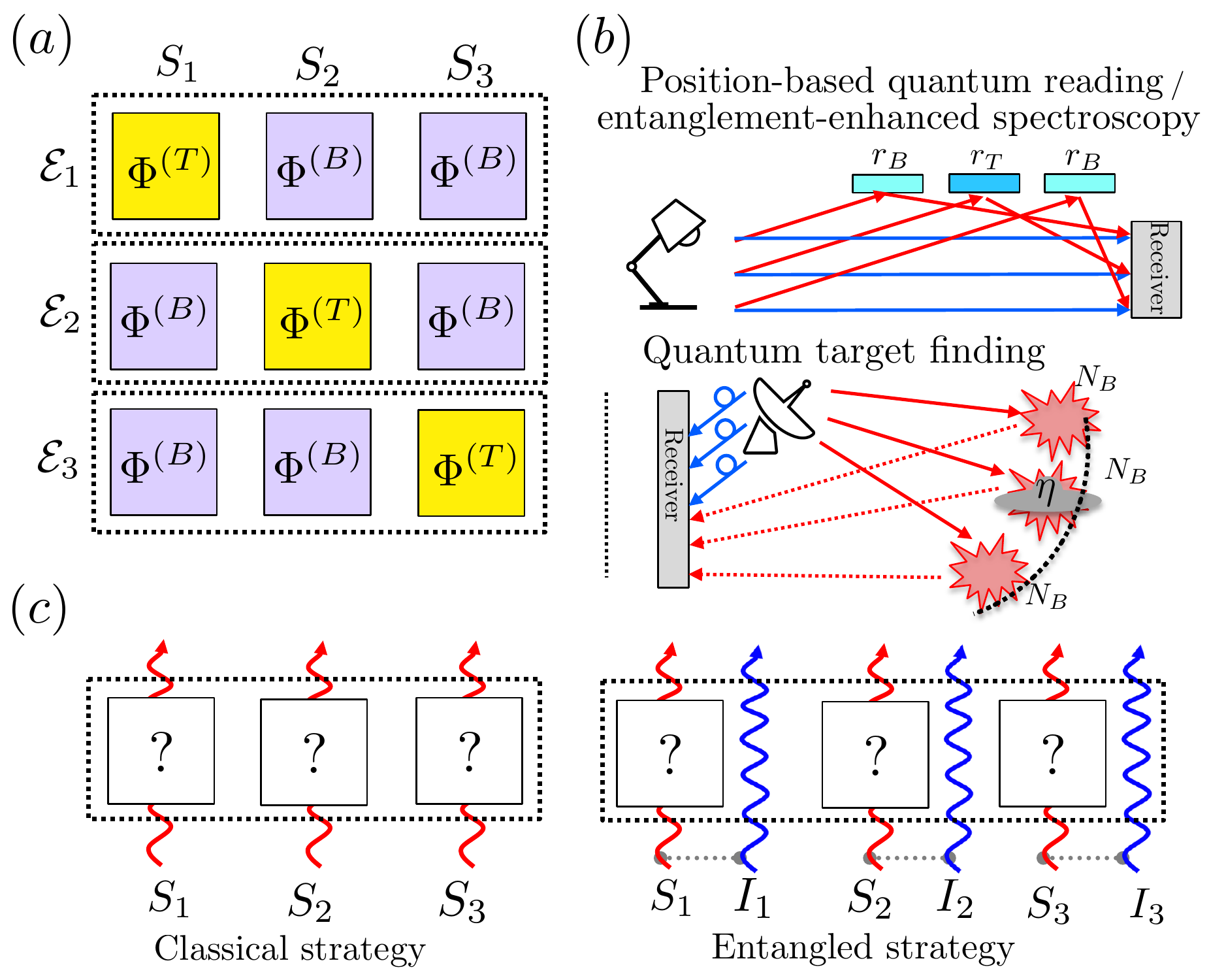}
\caption{{\bf Channel-position finding (CPF) schematics.} CPF represents a fundamental model of pattern recognition with quantum channels. (a) Example for $m=3$ \QZnew{subsystems}. Global channels $\calE_1,\calE_2,\calE_3$ consist of sub-channels $\Phi$ on subsystems $S_1,S_2,S_3$. Each sub-channel can be chosen to be a background channel $\Phi^{(B)}$ or a target channel $\Phi^{(T)}$. Channel $\calE_n$ \QZnew{(for $n=1,\cdots,m$)} means that the target channel is applied to subsystem $S_n$ while all the other subsystems undergo background channels.
(b) The classical strategy sends coherent-state signals (red, \QZnew{$S_k$}), while the entangled strategy sends signals (red, \QZnew{$S_k$}) entangled with locally stored idlers (blue, \QZnew{$I_k$}). (c) Bosonic applications to quantum reading of position-based data and quantum-enhanced direction finding of a remote target. Entangled pairs of signal (red) and idler (blue) are used. In position-based quantum reading, each sub-channel corresponds to a memory cell with reflectivity $r_B$ \SPnew{(background)} or $r_T$ \SPnew{(target)}; in quantum target finding, each sub-channel corresponds to a sector on a fixed-radius sphere where a target with reflectivity $\eta$ can be present or absent. If the target is absent, the returning signal is replaced by environmental noise with $N_B$ mean thermal photons per mode. 
\label{fig:schematic}
}
\end{figure}

\section{Results }

\subsection{General setting and main findings.}
We study the discrimination of multiple quantum channels by introducing and studying the problem of channel-position finding (CPF). This is a basic model of pattern recognition involving quantum channels, which has relations with the notion of pulse-position modulation~\cite{yuen1975optimum,sugiyama1989mppm,eldar2004optimal,cariolaro2010theory}. 
In CPF, a pattern is represented by a multi-mode quantum channel $\calE$ composed of $m$ sub-channels $\Phi$, each acting on a different subsystem $S_k$ (for $k=1,\ldots, m$) and chosen from a binary alphabet $\{\Phi^{(B)}, \Phi^{(T)}\}$. Only one of the sub-channels can be the target channel $\Phi^{(T)}$, while all the others are \SP{copies of a} background channel $\Phi^{(B)}$. A quantum pattern is therefore represented by a global channel $\calE_n$ \QZnew{(for $n=1,\cdots,m$)} where the target channel is only applied to subsystem $S_n$ while all the other subsystems undergo background channels (see Fig.~\ref{fig:schematic}a for \SPnew{a simple example with $m=3$).}

In this scenario, we design entanglement-enhanced protocols, based on a two-mode squeezed vacuum source and a generalized entangled version of the conditional-nulling (CN) receiver~\cite{dolinar1982near,dalla2014adaptive,chen2012optical,guha2011approaching}, that are able to greatly outperform any classical strategy based on coherent states (see Fig.~\ref{fig:schematic}\QZnew{b} for a schematic). This quantum advantage is quantified in terms of much lower mean error probability and improved error exponent for its asymptotic behavior.

Quantum-enhanced CPF has wide applications (see Fig.~\ref{fig:schematic}\QZnew{c}). In quantum reading of classical data, this corresponds to a novel formulation that we call `position-based quantum reading'. Here the information is encoded in the position of a target memory cell with reflectivity $r_T$ which is randomly located among background memory cells with reflectivity $r_B$. This is a particularly suitable model for information readout from sparse memory blocks. Changing from spatial to frequency modes, it can be mapped into a quantum-enhanced model of photometer or scanner, where the goal is to find an absorbance line within a band of frequencies. The advantage can therefore be interpreted as a quantum-enhanced tool for non-invasive spectroscopy. 

\SP{Another potential application of CPF is} quantum target finding, \SP{where} we simultaneously probe multiple space cells that are now represented by sectors of a sphere with some fixed radius. Only a single sector has a target with reflectivity $\eta$ while all the other sectors are empty. Moreover, each sector is characterized by bright noise so that $N_B$ mean thermal photons per bosonic mode are irradiated back to the receiver. Of course the problem is not limited to a spherical geometry. For instance, it can be seen in the context of defected device detection. Suppose there is an assembly line for producing a device that implements a channel, and with low probability, the assembly line produces a defective device that implements a different channel. Similarly, the problem can equivalently be mapped from spatial to frequency modes, so as to realize a quantum-enhanced scanner now working in very noisy conditions. 

\SP{Besides these potential applications, we expect that our results will have other implications beyond the model of CPF. For instance, as a by-product, we also found that our generalized CN receiver beats the best known receiver for the original binary problem of quantum reading~\cite{pirandola2011quantum} (see Sec.~\ref{me:receiver} for more details).}

\subsection{Generalized conditional nulling receiver}
From a mathematical point of view, the model of CPF exploits a relevant symmetry property that enables us to perform analytical calculations. Formally, we consider the discrimination of $m$ possible global channels $\{\calE_n\}_{n=1}^{m}$, each with equal prior probability and expressed by
\be
\calE_n=\big(\otimes_{k\neq n} \Phi^{(B)}_{S_k}\big)\otimes \Phi^{(T)}_{S_n},
\label{channel_GUS}
\ee 
where $\Phi^{(B/T)}_{S_k}$ is the background/target channel acting on subsystem $S_k$. In general, each subsystem may represent a collection of $M$ bosonic modes. 

It is easy to see that the ensemble of global channels $\{\calE_n\}_{n=1}^{m}$ has the geometric uniform symmetry (GUS)~\cite{cariolaro2010theory} 
$
\calE_n=S^{n-1} \calE_1 S^{\dagger n-1}
$,
where the unitary $S$ is a cyclic permutation and $S^m=I$, with $I$ being the identity operator. 
Because the channels are highly symmetric, it is natural to input a product state with GUS $\otimes_{k=1}^m\phi_{S_k}$, in which case the output state becomes
\be 
\rho_n=\big(\otimes_{k\neq n} \sigma^{(B)}_{S_k}\big) \otimes \sigma^{(T)}_{S_n},
\label{state_GUS}
\ee
where $\sigma^{(T/B)}:=\Phi^{(T/B)}(\phi)$. It is clear that this ensemble of output states also has GUS, i.e.,
$
\rho_n=S^{n-1} \rho_1 S^{\dagger n-1},
$
and it is analogous to the states considered in a pulse-position modulation~\cite{yuen1975optimum,eldar2004optimal,cariolaro2010theory}.

It is known~\cite{cariolaro2010theory,dalla2015optimality} that the optimal positive-valued operator measure (POVM) $\{\Pi_k\}$ minimizing the error probability for discriminating an ensemble of GUS states has the same type of symmetry, i.e., $\Pi_n=S^{n-1} \Pi_1 S^{\dagger n-1}$. This POVM has minimum error probability (Helstrom limit)
$
P_H=1-{\rm Tr}(\rho_1 \Pi_1).
$
For the specific cases where the output states are pure $\sigma^{T/B}=\ketbra{\psi^{(T/B)}}$, with overlap $\zeta=|\braket{\psi^{(T)}|\psi^{(B)}}|^2$, we have the following expression of the Helstrom limit  
\begin{align}
&P_H(m,\zeta)=\frac{m-1}{m^2}\left[\sqrt{1+(m-1)\zeta}-\sqrt{1-\zeta}\right]^2,
\label{P_H_PPM}
\end{align}
which is achievable by the `pretty good' measurement~\cite{PGM1,PGM2,PGM3}. In particular, note that for $m\zeta\ll1$ we have the asymptotic expansion
\be
P_H=\frac{1}{4}(m-1)\zeta^2+O(m^2\zeta^3).
\label{P_H_asym}
\ee

In general, when Eq.~(\ref{state_GUS}) represents an ensemble of mixed states, we do not know how to compute the ultimate Helstrom limit. However, we can resort to a sub-optimal detection strategy by generalizing the CN receiver of Ref.~\cite{dolinar1982near}. In fact, consider the $m$-ary CPF problem of Eq.~(\ref{channel_GUS}) with target/background channel $\Phi^{(T/B)}$. Assume that the pattern is probed by a GUS state so that the output ensemble is given by a generally-mixed state as in Eq.~(\ref{state_GUS}) with target/background state $\sigma^{(T/B)}$. Then, we show the following (\QZ{see Sec.~\ref{me:proof_theorem1} for a proof}).
\begin{theorem}[Generalized CN receiver]
\label{lemma:CN}
Denote by $h_n$ the hypothesis that the target channel $\Phi^{(T)}$ is encoded in sub-system $S_n$, so that the global channel is $\calE_n$. Suppose that there are two partially unambiguous POVMs, that we call t-POVM $\{\Pi_t^{(T)},\Pi_t^{(B)} \}$ and b-POVM $\{\Pi_b^{(T)},\Pi_b^{(B)} \}$, such that 
\be 
\tr[\Pi_t^{(T)} \sigma^{(T)}]= \tr[\Pi_b^{(B)} \sigma^{(B)}]=1.
\ee
Then, we design the following receiver. Start with $n=1$:

1. Check the current hypothesis $h_n$ by measuring subsystem $S_n$ with the t-POVM $\{\Pi_t^{(T)},\Pi_t^{(B)} \}$.

2. If the outcome from $S_n$ is `T', measure all the remaining subsystems $\{S_k\}_{k=n+1}^m$ in the b-POVM $\{\Pi_b^{(T)},\Pi_b^{(B)} \}$. If we get outcome `T' for some $S_k$ then select the hypothesis $h_k$. Otherwise, select $h_n$. 

3. If the outcome from $S_n$ is `B', then discard $h_n$ and repeat from point 1 with the replacement $n \rightarrow n+1$. If $n+1=m$, then select hypothesis $h_m$.

The error probability of this CN receiver is
\be
P_m^{\rm CN}(\zeta_1,\zeta_2)=\frac{1}{m}\frac{\zeta_2}{\zeta_1}\big(m\zeta_1+(1-\zeta_1)^m-1\big),
\label{P_CN}
\ee
where $\zeta_1=\tr(\sigma^{(B)}\Pi_t^{(T)})$ and $\zeta_2=\tr(\sigma^{(T)}\Pi_b^{(B)})$ are the two types of error probabilities.
\end{theorem}

\smallskip

\SP{Note that, when $m\zeta_1\ll1$, we have the asymptotic expansion
\be
P_m^{\rm CN}\simeq \frac{1}{2}(m-1)\zeta_1\zeta_2.
\ee
Also note that the above receiver is a CN receiver because it exploits partially-unambiguous POVMs and a feed-forward mechanism, similar to the classical CN receiver~\cite{dolinar1982near}. However, it is a \textit{generalized} CN receiver because it also involves entanglement with ancillas and may also be applied to mixed-state inputs, while the original CN receiver~\cite{dolinar1982near} only applies to pure states with no entanglement. Finally, our} receiver only relies on local operations and classical communication among the different subsystems, an important feature that makes it practical. 

For pure GUS states, one can always devise partially unambiguous POVMs and find symmetric error probabilities $\zeta_1=\zeta_2=\zeta$, in which case the CN receiver asymptotically achieves twice the Helstrom limit in Eq. (\ref{P_H_asym}). However, for mixed GUS states, it is generally difficult to design such POVMs, and we will have to give non-trivial constructions in this paper.
Also note that feed-forward is crucial for achieving good performance. 

In fact, suppose that we choose a simple strategy without feed-forward, e.g., measuring all subsystems in the $b$-POVM $\{\Pi_b^{(T)},\Pi_b^{(B)} \}$. In this case, no error occurs when measuring background states $\sigma^{(B)}$. \QZ{The error only occurs when this POVM is applied to the target state $\sigma^{(T)}$ and gives the erroneous outcome `B', which happens with probability $\zeta_2$. When this happens, we need to randomly guess (just because all outcomes would be equal to `B'). This gives a conditional error probability $\left(m-1\right)/m$, since only one among the $m$ subsystems is correct. The corresponding error probability for this design is given by
\be 
P_m^{\rm t}(\zeta_2)=\sum_{k=1}^m \frac{1}{m}\times \zeta_2\times \frac{m-1}{m} =(m-1)\zeta_2/m,
\label{Pmt}
\ee 
where the first $1/m$ factor is the equal prior.} 
We find that $P_m^{\rm t}(\zeta_2)\ge P_m^{\rm CN}(\zeta_1,\zeta_2)$, i.e., the CN strategy is always better than the non-feed-forward strategy and the advantage is particularly large when $\zeta_1$ is small.

\section{Classical versus entangled strategy}  
Given a CPF problem expressed by Eq.~(\ref{channel_GUS}), we aim to minimize the mean error probability affecting the discrimination of the corresponding $m$ hypotheses $\{h_n\}_{n=1}^{m}$. The solution of this problem is derived assuming that the signal modes irradiated over the subsystems are energetically-constrained. 
More precisely, let us discuss below the details on how we compare classical strategies (or `benchmarks') with quantum strategies.

In a classical strategy (see Fig.~\ref{fig:schematic}\QZnew{b}), we consider an input source which is described by a state with positive P-representation, so that it emits a statistical mixtures of multi-mode coherent states. First assume that this classical source has the GUS structure $\otimes_{k=1}^m\phi_{S_k}$, so that $M$ modes and $MN_S$ mean photons are irradiated over each subsystem. In this case, we can directly map Eq.~(\ref{channel_GUS}) into Eq.~(\ref{state_GUS}) and write the following lower bound based on Ref.~\cite{Barnum} (see Sec.~\ref{me:bounds} for more details)
\be 
P_{H,LB} = \frac{m-1}{2m}F^4\big(\sigma^{(T)},\sigma^{(B)}\big), \label{LB2_main}
\ee 
where $F$ is the quantum fidelity.

For the problem of CPF with arbitrary single-mode phase-insensitive bosonic Gaussian channels~\cite{holevo2007one,weedbrook2012gaussian} (see Sec.~\ref{me:channel} for a detailed definition), we prove a general classical benchmark. Suppose the target and background channels have transmissivity/gain $\mu_T$, $\mu_B$ and output noises $E_T$, $E_B$.
Given the most general classical source at the input, i.e., a multimode mixture of coherent states not necessarily with GUS structure, and assuming it irradiates a total of $mM$ modes and $mMN_S$ mean photons over the entire pattern of channels, we show the following lower bound \QZ{(LB)} to the mean error probability (see Sec.~\ref{App:classical_benchmarks} and Sec.~\ref{supp2} for proof) 
\be
P_{H,LB} = \frac{m-1}{2m} c_{E_B,E_T}^{2M}\exp\left[-\frac{2MN_S(\sqrt{\mu_B}-\sqrt{\mu_T})^2}{1+E_B+E_T} \right], 
\label{LB_Gaussi}
\ee
with $c_{E_B,E_T} \equiv [1+\big(\sqrt{E_B(1+E_T)}-\sqrt{E_T(1+E_B)}\big)^2]^{-1}$.

\QZ{First note that we can also obtain this bound from Eq.~(\ref{LB2_main}) by considering} a source that irradiates a single-mode coherent state $\ket{\sqrt{N_S}}$ for each of the $M$ modes probing subsystem $S_k$. Then, consider no passive signature $E_B=E_T$, which means that successful discrimination requires signal irradiation, i.e., it cannot be based on the passive detection of different levels of background noise. In this latter case, we find that an energetic single-mode coherent state $\ket{MN_S}$ on each subsystem is able to \QZ{produce Eq.~(\ref{LB_Gaussi}) from Eq.~(\ref{LB2_main})}. For this reason, in our next comparisons, we will also consider the performance of such a coherent-state source. In some cases, the \SP{corresponding output ensemble} will turn out to be pure, so that we can exactly quantify its performance via Eq.~(\ref{P_H_PPM}).

In order to obtain an enhancement by means of entanglement, we need to introduce ancillary `idler' systems $I_k$, for $1\le k \le m$, which are directly sent to the measurement apparatus (see Fig.~\ref{fig:schematic}\QZnew{b}). This means that the generic global channel takes the form
\be
\calE_n\otimes \calI=\Big[\otimes_{k\neq n} (\Phi^{(B)}_{S_k}\otimes \calI_{I_k})\Big]\otimes (\Phi^{(T)}_{S_n}\otimes \calI_{I_n}).
\ee
For the quantum source, we use the tensor product $\phi_{\rm ME}^{\otimes m M}$, where $\phi_{\rm ME}:=\sum_{k=0}^\infty \sqrt{N_S^k/(N_S+1)^{k+1}}\ket{k,k}$ is a two-mode squeezed vacuum state that maximally entangles a signal mode with a corresponding idler mode, given the mean number of photons $N_S$ constraining both signal and idler energies. Each subsystem $S_k$ is probed by the signal part of $\phi_{\rm ME}^{\otimes M}$ with a total of $MN_S$ photons on average irradiated over $S_k$. Therefore, the overall GUS ensemble of output states takes the form 
\be
\rho_n=(\calE_n \otimes I) \phi_{\rm ME}^{\otimes mM}=\big(\otimes_{k\neq n}\Xi^{(B)}_{S_kI_k} \big)\otimes \Xi^{(T)}_{S_n I_n},
\label{TMSV_out}
\ee 
where $\Xi^{(T/B)}=(\Phi^{(T/B)}\otimes \calI) (\phi_{ME}^{\otimes M})$. For generally-mixed states, it is difficult to calculate the Helstrom limit. One alternative is to use the upper bound (UB)~\cite{Barnum}
\be
 P_{H,UB} = (m-1)F^2\big(\Xi^{(T)},\Xi^{(B)}\big).
\label{UB2_main}
\ee
However, far better results can be found by employing the generalized CN receiver of Theorem~\ref{lemma:CN}. Note that the formulation and proof of this theorem automatically applies to the extended channel $\calE_n \rightarrow \calE_n\otimes \calI$ and the corresponding target/background state $\sigma^{(T/B)} \rightarrow \Xi^{(T/B)}$.

In the following we explicitly compare classical and quantum performance for the paradigmatic cases mentioned in our introduction, i.e., position-based quantum reading and quantum target finding, including their frequency-based spectroscopic formulations. In all cases we exactly quantify the quantum advantage that is achievable by the use of entanglement.

\section{Position-based quantum reading and frequency scanner} 
As depicted in Fig.~\ref{fig:schematic}, a possible specification of the problem is for the quantum readout of classical data from optical memories. In quantum reading~\cite{pirandola2011quantum}, the bosonic channels are used to model the reflection of light from the surfaces of an optical cell with different reflectivities, whose two possible values $r_T$ and $r_B$ are used to encode a classical bit. In the absence of other noise, the readout process is therefore equivalent to discriminating the value $r \in \{r_T,r_B\}$ of the loss parameter of a pure-loss bosonic channel $\calL_{r}$. In our position-based formulation of the protocol, the classical information is encoded in the position of a target cell (with reflectivity $\mu_T=r_T$) within a pattern of $m$ cells, where all the remaining are background cells (with reflectivity $\mu_B=r_B$). In general, we probe each cell with $M$ bosonic modes, so that we have target channel $\Phi^{(T)}=\calL_{r_T}^{\otimes M}$ and background channel $\Phi^{(B)}=\calL_{r_B}^{\otimes M}$. In the following, we develop our theory of position-based quantum and classical reading in this pure-loss setting, where $E_B=E_T=0$. Our analysis can be extended to the presence of extra noise (thermal-loss channels) as discussed in Sec.~\ref{noise_reading}.

As previously mentioned, we can map the model from spatial to frequency modes. This means that the problem may be translated into a spectroscopic one where the goal is to find a faint absorbance line $r_T<1$ within a range $W$ of transparent frequencies ($r_B\sim 1$). This can be resolved into a discrete ensemble of $m=W/\delta W$ modes, where $\delta W$ is the bandwidth of the detector. The corresponding quantum-advantage can then be directly re-stated in terms of better identifying an absorbance line in a frequency spectrum, where we are constrained to use a white power spectral density over $W$ for a certain time duration, so that the total irradiated energy is equal to $m MN_S$. This model can be considered both in transmission (e.g., in a spectro-photometer setup) and in reflection (e.g., in a scanner-like setup).

\subsection{Position-based reading with classical light} 
We can easily specify the lower bound in Eq.~(\ref{LB_Gaussi}) to the reading problem, so that we get the following lower bound for position-based classical reading of a block of $m$ cells irradiated by $mMN_s$ mean photons
\be 
P_{H,LB}^{\rm CR}= \frac{m-1}{2m}e^{-2M N_S (\sqrt{r_B}-\sqrt{r_T})^2},
\label{LB_CR}
\ee 
\QZ{where `CR' stands for classical reading.}
As discussed before, \QZ{we can also obtain this bound from Eq.~(\ref{LB2_main}) by irradiating} energetic single-mode coherent states on each subsystem, i.e., $\otimes_{k=1}^m \ket{\alpha}_{S_k}$ with $\alpha=\sqrt{MN_S}$. 

Assuming the input source $\otimes_{k=1}^m \ket{\alpha}_{S_k}$, the output states $\{ \rho_n \}_{n=1}^m$ are pure, expressed by Eq.~(\ref{state_GUS}) with $\sigma^{(\ell)}=\ket{\sqrt{r_\ell}\alpha}$ for $\ell=T,B$. Thus we can use Eq.~(\ref{P_H_PPM}) to calculate the Helstrom limit at the output
\be
P_{H}^{\rm CR}(r_B,r_T,M,N_S)=P_H(m,\zeta^{\rm CR}),
\label{P_H_QR_C}
\ee
where 
$ 
\zeta^{\rm CR}=|\braket{\sqrt{r_B}\alpha|\sqrt{r_T}\alpha}|^2=e^{-M N_S (\sqrt{r_B}-\sqrt{r_T})^2}
$.
In the limit of small overlap $\zeta\ll1$, we have 
\be 
P_{H}^{\rm CR}\simeq \frac{1}{4}(m-1)e^{-2M N_S (\sqrt{r_B}-\sqrt{r_T})^2},
\label{P_H_QR_Classical_asym}
\ee 
which is only $m/2$ times larger than the lower bound in Eq.~(\ref{LB_CR}). This also means that the lower bound is tight in the error exponent. Although it is extremely difficult to minimize the Helstrom limit by varying the input among general non-symmetric classical states, we can show that mixtures of the type $\int d^2\alpha P(\alpha) \otimes_{k=1}^m \ket{\alpha}_{S_k}$ or increasing the modes in each subsystem do not improve the value of $P_{H}^{\rm CR}$ (see details in Sec.~\ref{me:pure}).


\subsection{Position-based reading with entangled light}
To get a quantum advantage in terms of a lower error probability and, therefore, a higher rate of data retrieval from the pattern, we interrogate each cell with the signal-part of an $M$-pair two-mode squeezed vacuum state $\phi_{\rm ME}^{\otimes M}$. At the output of each cell, we get the state $\Xi^{(\ell)}=\left[(\calL_{r_\ell}\otimes \calI) \phi_{\rm ME}\right]^{\otimes M}$ for $\ell=B,T$. We can upper bound the error probability using the formula in Eq.~(\ref{UB2_main}), where the fidelity term $F^2\big(\Xi^{(T)},\Xi^{(B)}\big)=F^{2M}\big[(\calL_{r_T}\otimes \calI) \phi_{\rm ME},(\calL_{r_B}\otimes \calI) \phi_{\rm ME}\big]$ can be exactly calculated (see Sec.~\ref{supp1} for details). The exact expression of the bound $P_{H,UB}^{\rm QR}$ is too long to display, but will be used in our numerical comparisons (\QZ{here `QR' stands for quantum reading}). 

For $N_S\ll1$ and $M\gg1$ at fixed $MN_S$ per cell, we have the simple asymptotic expansion
\be 
P_{H,UB}^{\rm QR}
\simeq (m-1)e^{-2MN_S(1-\sqrt{(1-r_B)(1-r_T)}-\sqrt{r_Br_T})}.
\label{P_H_QR_E_UB_asym}
\ee 
Comparing Eqs.~(\ref{P_H_QR_Classical_asym}) and~(\ref{P_H_QR_E_UB_asym}), we can already see that, for $r_T+r_B\ge 1$, the error exponent of the quantum case is better than the exact error exponent of the classical case. In particular, this advantage becomes large when both $r_T$ and $r_B$ are close to unity. 

We can improve this result and show a greater quantum advantage by employing the generalized CN receiver of Theorem~\ref{lemma:CN}. An important preliminary observation is that the output state $(\calL_{r}\otimes \calI) \phi_{\rm ME}$, from each probing of a generic cell, can be transformed into a tensor product form, where the signal mode is in the vacuum state and the idler mode is in a thermal state with mean photon number $(1-r)N_S$. This is possible by applying a two-mode squeezing operation $S_2[s(r,N_S)]$, with strength
\be
s(r,N_S)=\frac{1}{2}\ln\left(\frac{\sqrt{N_S+1}-\sqrt{r N_S}}{\sqrt{N_S+1}+\sqrt{r N_S}}\right).
\ee

This allows us to design a CN receiver for the cell output state $\Xi^{(\ell)}$, which consists of two-mode squeezing operations followed by photon counting on the signal modes. By applying $S_2[s(r_B,N_S)]$ to each pair of the $2M$ signal-idler modes, we have that $\Xi^{(B)}$ is transformed into a state $\tilde{\Xi}^{(B)}$ with vacuum signal modes; while $\Xi^{(T)}$ becomes a state $\tilde{\Xi}^{(T)}$ where the signal modes are in a product of $M$ thermal states, each with mean photon number 
\be
n(N_S,r_B,r_T)=\frac{N_S(N_S+1)(\sqrt{r_B}-\sqrt{r_T})^2}{1+N_S(1-r_B)}.
\ee

Let us now measure the number of photons on the $M$ signal modes. The outcomes are interpreted as follows: If we count any photon then return `T', otherwise return `B'. Assuming this rule, the background state $\tilde{\Xi}^{(B)}$ does not lead to any photon count and, therefore, to any error. An error occurs only if, in the presence of a target state $\tilde{\Xi}^{(T)}$, we get zero count on all $M$ signal modes, which happens with probability
\begin{equation}
\zeta_2^{\rm QR}=[1+n(N_S,r_B,r_T)]^{-M}.\label{QRM1}
\end{equation}
\SP{This measurement} implements the $b$-POVM of our CN receiver (unambiguous over background cells).

\begin{figure}
\vspace{-0.0cm}
\centering\includegraphics[width=0.49\textwidth]{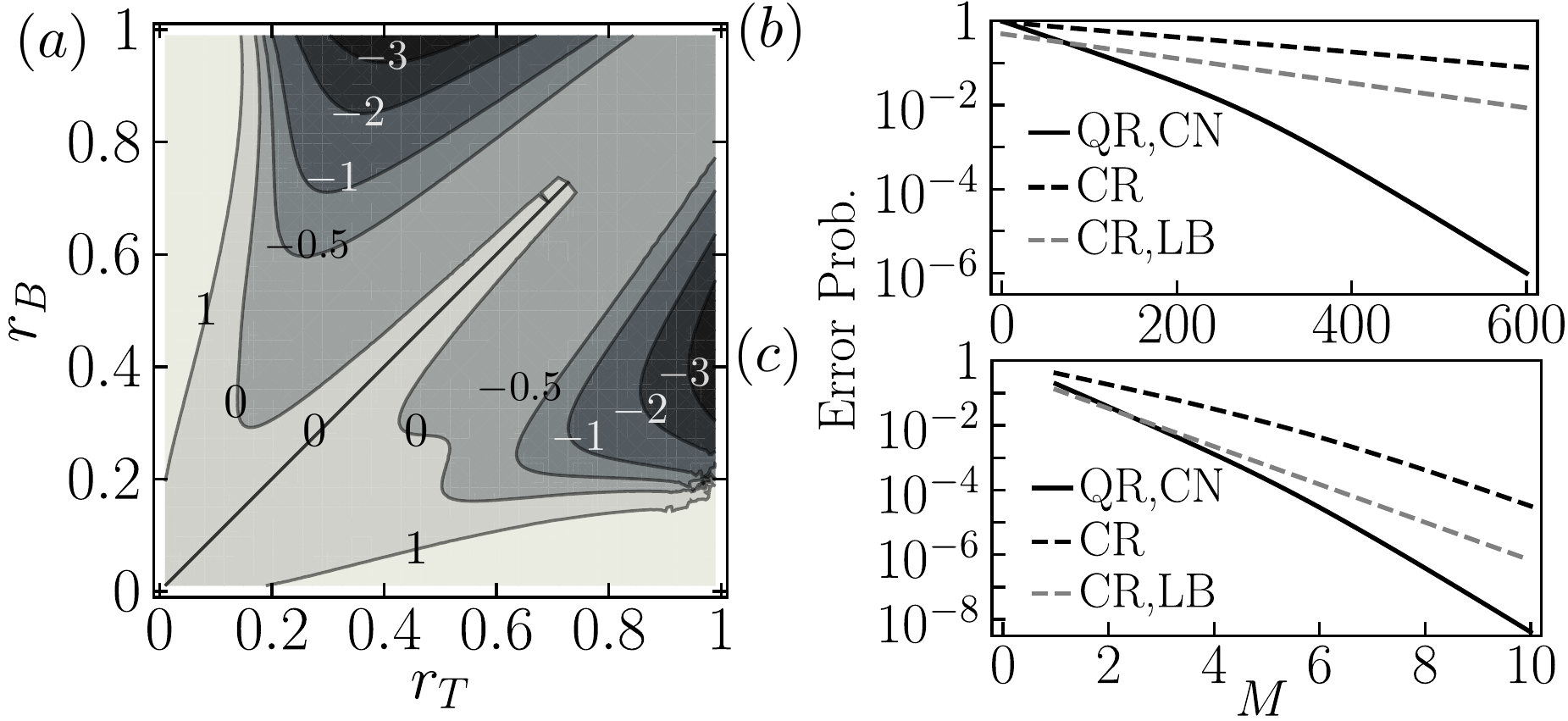}
\caption{{\bf Position-based quantum reading.} Quantum advantage shown for a block of $m=100$ cells and $N_S=5$ mean photons per mode. (a) 
\QZnew{We consider the log ratio of the error probabilities ($\log_{10}[P_{CN}^{\rm QR}/P_{H}^{\rm CR}]$), between quantum reading with conditional-nulling receiver $P_{CN}^{\rm QR}$ and classical reading in the Helstrom limit $P_{H}^{\rm CR}$. This ratio is plotted as a function of the background and target reflectivities, $r_B$ and $r_T$, for $M=10$ modes per cell.} Note that since Eq.~(\ref{P_CN}) is not symmetric in $r_B$ and $r_T$, we observe asymmetric patterns.
(b) Error probabilities $P_{CN}^{\rm QR}$ (black solid) and $P_{H}^{\rm CR}$ (black dashed) versus number of modes $M$, for reflectivities $r_B=0.95$ and $r_T=0.9$. We also include the \SPnew{ultimate} classical benchmark given by \QZnew{the lower bound for classical reading} $P_{H,LB}^{\rm CR}$ (gray dashed). (c) As in panel~(b) but with $r_B=1$ and $r_T=0.4$.
\label{fig:PCN_PE}
}
\end{figure}

Let us now realize the $t$-POVM, which is unambiguous on target cells. In this case, we apply the operator $S_2[s(r_T,N_S)]$ with different squeezing, so that $\tilde{\Xi}^{(T)}$ has vacuum signal modes, while $\tilde{\Xi}^{(B)}$ has thermal signal modes, each with mean photon number $n(N_S,r_T,r_B)$. By performing photon counting on the signal modes and using the same rule above, we have that an error occurs only if a background state $\tilde{\Xi}^{(B)}$ gets zero counts on all $M$ modes, which happens with probability
\be
\zeta_1^{\rm QR}=[1+n(N_S,r_T,r_B)]^{-M}.
\ee

We can now study the performance of the CN receiver from Theorem~\ref{lemma:CN}, where we use the formula of Eq.~(\ref{P_CN}) computed over the two types of error probabilities $\zeta_1^{\rm QR}$ and $\zeta_2^{\rm QR}$. For position-based quantum reading of a block of $m$ cells, we find the achievable error probability
\be
P_{CN}^{\rm QR}=P_m^{\rm CN}(\zeta_1^{\rm QR},\zeta_2^{\rm QR}).
\label{P_CN_QR_G}
\ee
At low photon numbers $N_S\ll1$ while keeping the total irradiated energy $MN_S$ as a finite value, we have that $P_{CN}^{\rm QR}\simeq 2 P_{H}^{\rm CR}(r_B,r_T,M,N_S)$, i.e., a factor of two worse than the classical performance in Eq.~(\ref{P_H_QR_Classical_asym}). However, for larger values of $N_S$ and assuming the condition $N_S(\sqrt{r_B}-\sqrt{r_T})^2\ll1$, we find that  
\be
P_{CN}^{\rm QR}\simeq \frac{m-1}{2} e^{-M(N_S+1)(\sqrt{r_B}-\sqrt{r_T})^2\big(\frac{1}{1-r_T}+\frac{1}{1-r_B}\big)},
\ee
which has a large advantage in the error exponent when $r_B$ and $r_T$ are close to $1$, as also evident from Fig.~\ref{fig:PCN_PE}.

\subsection{Further quantum enhancement} 
Let us consider an ideal scenario for position-based quantum reading, where the target cell with $r_T<1$ has to be found among many background cells with perfect reflectivity $r_B=1$. This configuration allows us to show an even higher quantum advantage. In fact, for ideal background ($r_B=1$), the application of $S_2[s(r_B,N_S)]$ generates a background state $\tilde{\Xi}^{(B)}$ which is vacuum in all signal and idler modes, and a target state $\tilde{\Xi}^{(T)}$ which is non-vacuum on all these modes. We can therefore apply the $b$-POVM of the CN receiver to the entire set of $2M$ signal and idler modes.

\begin{figure}
\vspace{-0cm}
\centering\includegraphics[width=0.49\textwidth]{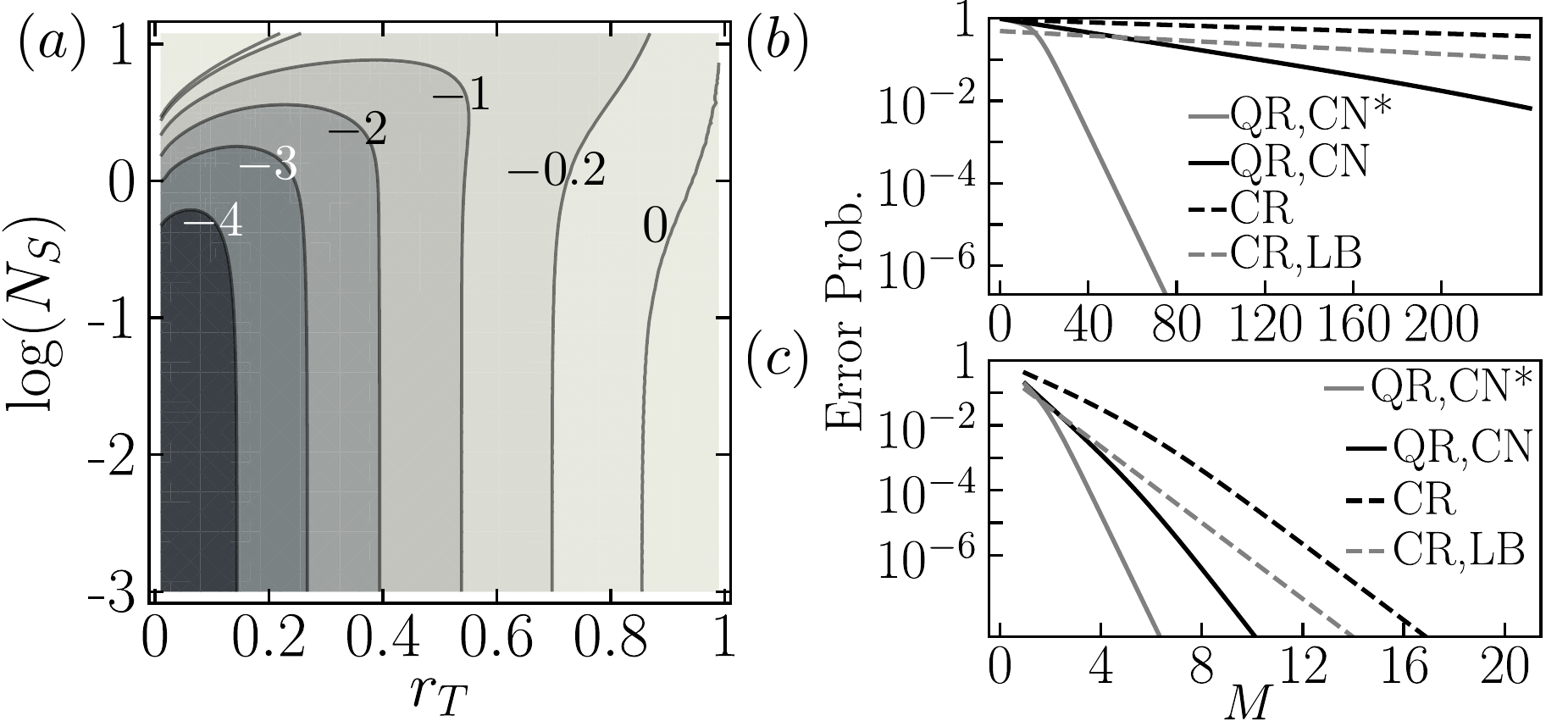}
\caption{{\bf Position-based quantum reading \QZnew{with ideal background}.} Quantum advantage for ideal background \QZnew{reflectivity} ($r_B=1$) and considering $m=100$ cells. (a) 
\QZnew{We consider the log ratio of the error probabilities ($\log_{10}[P_{CN*}^{\rm QR}/P_{H}^{\rm CR}]$), between quantum reading with improved conditional-nulling receiver $P_{CN*}^{\rm QR}$ and classical reading in the Helstrom limit $P_{H}^{\rm CR}$. This ratio is plotted} as a function of the target reflectivity $r_T$ and mean photon number per mode $N_S$ for fixed $MN_S=12$, \QZnew{where $M$ is the number of modes}. (b) We show the various error probabilities, i.e., \QZnew{quantum reading with the improved conditional nulling receiver} $P_{CN*}^{\rm QR}$ (including measurements of the idlers, gray solid), \QZnew{quantum reading with the conditional nulling receiver} $P_{CN}^{\rm QR}$ (based on the measurement of the signals only, black solid), the classical performance $P_{H}^{\rm CR}$ (black dashed), and the ultimate classical benchmark $P_{H,LB}^{\rm CR}$ (gray dashed). These are plotted versus the number of modes $M$, for $r_T=0.95$ and $N_S=5$. (c) As in panel~(b) but choosing parameters $r_T=0.4$ and $N_S=5$.
\label{fig:PCN_PE_rb1}
}
\end{figure}

The type-II error probability is obtained by calculating the fidelity between $\tilde{\Xi}^{(T)}$ and the vacuum state (see Sec.~\ref{supp1} for details). This leads to
\be
\zeta_{2*}^{\rm QR}
=\big[1+N_S(1-\sqrt{r_T})\big]^{-2M} 
=\frac{\zeta_2^{\rm QR}}{[1+N_S(1-r_T)]^M},\label{QRM2}
\ee
with a clear improvement with respect to the previous case $\zeta_2^{\rm QR}$. Consider now the $t$-POVM. The application of the other squeezing operator $S_2[s(r_T,N_S)]$ generates a target state $\tilde{\Xi}^{(T)}$ with vacuum signals but non-vacuum idlers, so that we must again restrict photon counting to the signal modes, implying that we achieve the same type-I error probability as before, i.e., $\zeta_{1*}^{\rm QS}=\zeta_1^{\rm QR}$. 

Using Eq.~(\ref{P_CN}), we derive the overall error probability $P_{CN*}^{\rm QR}=P_m^{\rm CN}(\zeta_{1*}^{\rm QR},\zeta_{2*}^{\rm QR})$. At low photon numbers $N_S\ll1$ while keeping the total energy $MN_S$ as finite, we find
\be 
P_{CN*}^{\rm QR}\simeq P_{H}^{\rm CR}(1,r_T,M,N_S)\times 2e^{-MN_S(1-r_T)},
\label{PCN_star}
\ee 
which shows a large advantage in the error exponent with respect to the classical strategy of Eq.~(\ref{P_H_QR_Classical_asym}). In Fig.~\ref{fig:PCN_PE_rb1} we show the quantum advantage both in terms of error exponent and actual values of the error probabilities. This further quantum enhancement is particularly relevant to spectroscopy, where the background is indeed highly transparent with $r_B$ very close to unity.

Finally, let us note that the other case of $r_T=1$ and $r_B<1$ can be improved in the same way, leading to an improved type-I error probability
\be
\zeta_{1*}^{\rm QR}
=\big[1+N_S(1-\sqrt{r_B})\big]^{-2M} 
=\frac{\zeta_1^{\rm QR}}{[1+N_S(1-r_B)]^M},
\ee
and the overall error probability 
\be 
P_{CN*}^{\rm QR}\simeq P_{H}^{\rm CR}(r_B,1,M,N_S)\times 2e^{-MN_S(1-r_B)}.
\ee

\section{Quantum target finding} 
In general, target detection involves a search in multiple space-time-frequency bins. Time bins are associated with ranging, frequency bins can be used for speed detection via Doppler effect, while space bins are associated with direction finding. Let us study the latter problem here, i.e., discovering the position of a single target in terms of polar and azimuthal angles, while we assume it is at some fixed range $R$ and does not create large Doppler shifts. Let us divide the $R$-radius horizon sphere into $m$ non-overlapping sectors, one of which contains the reflective target. For large $m$, each sector $S_k$ is approximately subtended by a corresponding small solid angle (see Fig.~\ref{fig:schematic}).

We simultaneously probe all $m$ sectors, while using $M$ bosonic modes for each of them (e.g., a train of temporal pulses or a single broadband pulse). Each signal mode will shine $N_S$ mean number of photons. Let us denote by $\mathcal{L}_\mu^{N}$ a thermal-loss channel with loss parameter $\mu$ and mean number of thermal photons $N$, so that its output noise is $E=(1-\mu)N$. When the target is present in a sector, the $M$ signal modes go through the target channel $\Phi^{(T)}={\big(\mathcal{L}_\eta^{N_B/(1-\eta)}\big)}^{\otimes M}$, so that each mode is affected by loss parameter $\mu_T=\eta$ and output noise $E_T=N_B$. 
By contrast, if the target is absent in a sector, then the $M$ signal modes are lost and replaced by environmental modes, each having $N_B$ mean thermal photons. For target absent, we therefore have the background channel $\Phi^{(B)}={\big(\mathcal{L}_0^{N_B}\big)}^{\otimes M}$, with $\mu_B=0$ and $E_B=N_B$ (no passive signature). 

\QZ{We consider the region of quantum illumination~\cite{tan2008quantum}, where bright thermal noise $N_B\gg1$ is present in the environment, as it would be the case at the microwave wavelengths~\cite{ShabirPRL}. We then consider low energy signals ($N_S\ll1$) so that the probing is non-revealing and/or non-destructive for the target. In these conditions, the considered quantum channels are clearly entanglement-breaking. Before we present the corresponding results, let us note that the model for target finding can also be mapped to a model of quantum-enhanced frequency scanner, now in the presence of bright environmental noise. See Sec.~\ref{me:ranging} for more details on this mapping and also for a discussion on target ranging.}

\subsection{Target finding with classical light}
The general lower bound in Eq.~(\ref{LB_Gaussi}) can be specified to classical target finding, by setting $E_T=E_B=N_B$ and $\mu_T=\eta, \mu_B=0$, so that we have 
\be 
P_{H,LB}^{\rm CTF}=\frac{m-1}{2m}
\exp\left[-\frac{2M\eta N_S}{2N_B+1}\right],
\label{P_H_CI_LB}
\ee 
\QZ{where `CTF' stands for classical target finding}.
This expression bounds the best performance achievable by classical sources of light that globally irradiate $mMN_S$ mean photons over the entire sphere. \QZ{In particular, we can also obtain this bound from Eq.~(\ref{LB2_main}) by considering} $m$ single-mode coherent states $\otimes_{k=1}^m \ket{\sqrt{MN_S}}_{S_k}$, each shining $M N_S$ mean photons on a sector. 

Let us compute the classical performance with a specific receiver. When we use the uniform coherent source $\otimes_{k=1}^m \ket{\sqrt{MN_S}}_{S_k}$ at the input, the ensemble of output states of Eq.~(\ref{state_GUS}) is defined on the following background and target states
\begin{align}
& \sigma^{(B)}=\mathcal{L}_0^{N_B}\left(\ketbra{\sqrt{MN_S}}\right), \\
& \sigma^{(T)}=\mathcal{L}_\eta^{N_B/(1-\eta)}\left(\ketbra{\sqrt{MN_S}}\right). 
\end{align}
This is identical to classical pulse-position modulation decoding with signal $\sqrt{\eta MN_S}$ and thermal noise $N_B$~\cite{cariolaro2010theory}.
We can therefore consider the direct detection (DD) scheme based on photon counting \QZ{(see Ref.~\cite[p.~193]{Helstrom_1976} and Ref.~\cite{cariolaro2010theory})}, giving the error probability
\begin{align}
P_{DD}^{\rm CTF}=&\frac{1}{m}\sum_{k=2}^m (-1)^k C_m^k \cross \nonumber \\ 
& \exp\left[-\frac{(1-v)(1-v^{k-1})\eta MN_S}{1-v^k}\right],
\label{P_CI_DD}
\end{align}
where $v=N_B/(N_B+1)$ and $C_m^k$ is the binomial coefficient (number of combinations of $k$ items out of $m$). 

In the high-noise $N_B\gg1$ and large number of modes $M\gg1$ limit, this error probability is dominated by the smallest error exponent in the sum, and it becomes 
\be
P_{DD}^{\rm CTF}\simeq \frac{m-1}{2m}\exp\left(-M\eta N_S/2N_B\right).\label{DD_CC_asy}
\ee
This is only a factor $2$ worse than the bound in Eq.~(\ref{P_H_CI_LB}). In these limits, we expect that classical target finding via a DD scheme is close to the optimum.

\begin{figure}
\centering
\includegraphics[width=0.35\textwidth]{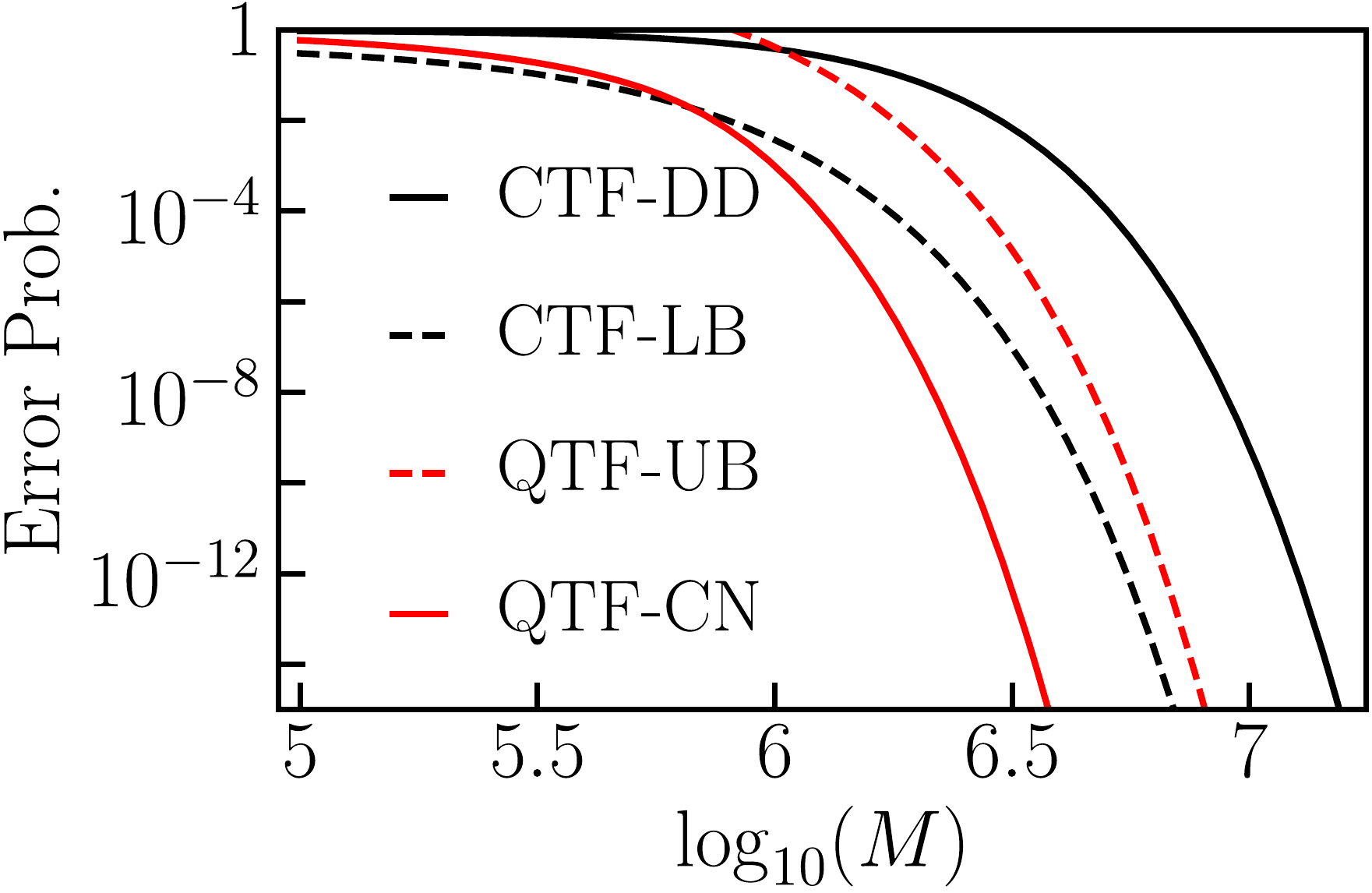}
\caption{{\bf Target direction finding with classical and entangled light.} We plot the error probabilities in terms of number of modes $M$, considering $m=50$ sectors, $N_S=10^{-3}$ photons per mode, $N_B=20$ thermal photons per environmental mode, and $\eta=0.1$ round-trip loss. We consider the performance of classical target finding via direct detection from Eq.~(\ref{P_CI_DD}) (CTF-DD, solid black line) and assuming the lower bound of Eq.~(\ref{P_H_CI_LB}) (CTF-LB,  black dashed line). We then consider the performance of quantum target finding assuming the upper bound of Eq.~(\ref{P_H_QI_E_UB_asym}) (QTF-UB, red dashed line) and via the generalized CN receiver from Eq.~(\ref{P_CN_QI}) (QTF-CN, solid red line).  
\label{fig:PCN_QI}
}
\end{figure}

\subsection{Target finding with entangled light} 
Let us now assume a tensor product of two-mode squeezed vacuum states $\phi_{\rm ME}^{\otimes mM}$ at the input. In each $M$-mode probing of a sector, the ensemble of possible output states takes the form of Eq.~(\ref{TMSV_out}) with the following background and target states
\begin{align}
& \Xi^{(B)}={\big[(\mathcal{L}_0^{N_B}\otimes \calI) \phi_{\rm ME}\big]}^{\otimes M}, \\
&\Xi^{(T)}={\big[(\mathcal{L}_\eta^{N_B/(1-\eta)}\otimes \calI) \phi_{\rm ME}\big]}^{\otimes M}.
\end{align}
Let us compute an upper bound based on Eq.~(\ref{UB2_main}). Its exact expression is too long to display, even though it is used in our numerical evaluation. In the limits of $N_S\ll1$ and $M\gg1$ while keeping the total energy per sector $MN_S$ as fixed, we find the following asymptotic bound for quantum target finding
\be 
P_{H,UB}^{\rm QTF}(\eta,N_B,M,N_S) \simeq (m-1)\exp\left(-\frac{M\eta N_S}{1+N_B}\right),
\label{P_H_QI_E_UB_asym}
\ee
\QZ{where `QTF' stands for quantum target finding.}
This has no advantage with respect to Eq.~(\ref{P_H_CI_LB}), but both bounds are likely to be non-tight. It has instead a factor of $2$ advantange in the error exponent with respect to the direct detection result in Eq.~(\ref{DD_CC_asy}) for large noise. To better evaluate the performance of the entangled case, we need to analyze an explicit  receiver design. 

We adapt the quantum illumination receiver based on sum-frequency-generation (SFG) process~\cite{zhuang2017optimum} to the CN approach in Theorem~\ref{lemma:CN}. Consider the problem of binary hypothesis testing between the states $\Xi^{(B)}$ and $\Xi^{(T)}$. An SFG receiver converts the signal-idler cross correlations into photon number counts, through the combination of multiple cycles of SFG process and interference. In the limit of $N_S\ll1$ and $N_B\gg1$ with feed-forward disabled, the photon counting statistics of $\Xi^{(T)}$ is equivalent to a coherent state with mean photon number $M\eta N_S(N_S+1)/N_B$, and $\Xi^{(B)}$ is equivalent to a vacuum state. 

After this conversion, suppose we perform the photon-counting stage of the SFG measurement on the background state $\Xi^{(B)}$, then there is always zero count and therefore no ambiguity. For $\Xi^{(T)}$, there is instead some type-II probability $\zeta_2^{\rm QTF}=e^{-M\eta N_S(N_S+1)/N_B}$ of getting zero count and therefore selecting the wrong hypothesis `B'. This corresponds to the $b$-POVM of the generalized CN receiver. On the other hand, for the $t$-POVM, suppose we apply a two-mode squeezer $S_2(r^{\rm QTF})$ before performing the previous SFG measurement, where
\be
r^{\rm QTF}=-\frac{1}{2}\arctan\left[\frac{-2\sqrt{\eta N_S(N_S+1)}}{1+N_S+N_B}\right]
\ee
is chosen such that $S_2(r^{\rm QTF})\Xi^{(T)} S_2^\dagger(r^{\rm QTF})$ has zero cross correlations. Then we decide `T' when no photon is counted, making no error. However, when the input is $\Xi^{(B)}$, the squeezer will create phase sensitive cross correlations $\simeq \sqrt{\eta N_S(N_S+1)}$. When no counts are registered, we select the wrong hypothesis `T', with type-I error probability $\zeta_1^{\rm QTF}=\zeta_2^{\rm QTF}$. 

According to Theorem~\ref{lemma:CN}, the performance of the generalized CN receiver (here applied to signals and idlers) corresponds to the following mean error probability
\begin{align}
P_{CN}^{\rm QTF}=P_m^{\rm CN}(\zeta_1^{\rm QTF},\zeta_2^{\rm QTF})
\simeq \frac{1}{2}(m-1)e^{-2M\eta N_S/N_B}.
\label{P_CN_QI}
\end{align}
Comparing with Eq.~(\ref{P_H_CI_LB}), we see that the achievable performance of quantum target finding clearly outperforms the bound on classical target finding. In particular, we see that the error exponent is increased by a factor $2$. We explicitly compare these results in Fig.~\ref{fig:PCN_QI}.

\section{\bf Discussion}

In this work we showed that the use of quantum entanglement can remarkably enhance the discrimination of multiple quantum hypotheses, represented by different quantum channels. More precisely, we considered a basic problem of quantum pattern recognition that we called channel-position finding. \SPnew{This model can also be regarded as a quantum channel formulation of the classical notion of pulse position modulation~\cite{PPMref}, so that it clearly departs from other approaches that exploit pulse position modulation for \textit{state-based} encoding (e.g.,~\cite{warsi}).} 
In this scenario, we showed that the use of an entangled source and a suitably constructed conditional-nulling receiver can outperform any classical strategy in finding the unknown position of the  channel. This quantum advantage, which is quantified in terms of improved error probability and error exponent, has been demonstrated for paradigmatic examples of position-based quantum reading and quantum target finding, besides their spectroscopic formulations as quantum-enhanced frequency scanners. \SP{As further theoretical directions, it would be interesting to exactly establish the optimal performance for discriminating quantum channels with geometrical uniform symmetry.} \QZ{Finally, although our analysis relies on symmetry, we expect that a similar quantum advantage exists in problems with completely arbitrary channel patterns.}

\section{Methods}

\subsection{Phase-insensitive bosonic Gaussian channels} 
\label{me:channel}
\QZ{
The action of a single-mode (covariant) phase-insensitive Gaussian channel over input quadratures $\hat{\bm x}=(\hat{q},\hat{p})^T$ can be represented by the transformation $\hat{\bm x} \rightarrow \sqrt{\mu} \hat{\bm x} + \sqrt{|1-\mu|}\hat{\bm x}_{E} + \xi$, where $\mu$ is a transmissivity ($0 \le \mu \le 1$) or a gain ($\mu \ge 1$), $\hat{\bm x}_{E}$ are the quadratures of an environmental mode in a thermal state with noise variance $\omega = 2N+1$ with $N$ being the mean number of photons, and $\xi$ is additive classical noise, i.e., a random 2D Gaussian distributed vector with covariance matrix $w_{\rm add}\mathbf{I}$. Here we assume vacuum shot noise equal to $1$.}

\QZ{
Note that, for a coherent state at the input, the output state of the channel is generally thermal with covariance matrix $\mathbf{V} = (\mu + |1-\mu|\omega + \omega_{\rm add})\mathbf{I}$. Setting $\omega = (1+2E-\omega_{\rm add}-\mu)/|1-\mu|$, this matrix simply becomes $(2E+1)\bm I$. Therefore, conditionally on a coherent state input, the channel can be described by the two parameters $\mu$ and $E$. In particular, for a thermal-loss channel, we have $1 \le \mu \le 1$, and $E=(\omega-1)(1-\mu)/2=(1-\mu)N$; for a noisy amplifier, we have $\mu \ge 1$, and $E=(\omega+1)(\mu-1)/2=(\mu-1)(N+1)$; and finally, for an additive Gaussian noise channel, we have $\mu = 1$ and $E=\omega_{\rm add}/2$.
}

\subsection{Optimal receiver design for standard quantum reading}
\label{me:receiver}
\begin{figure}
\centering
\includegraphics[width=0.49\textwidth]{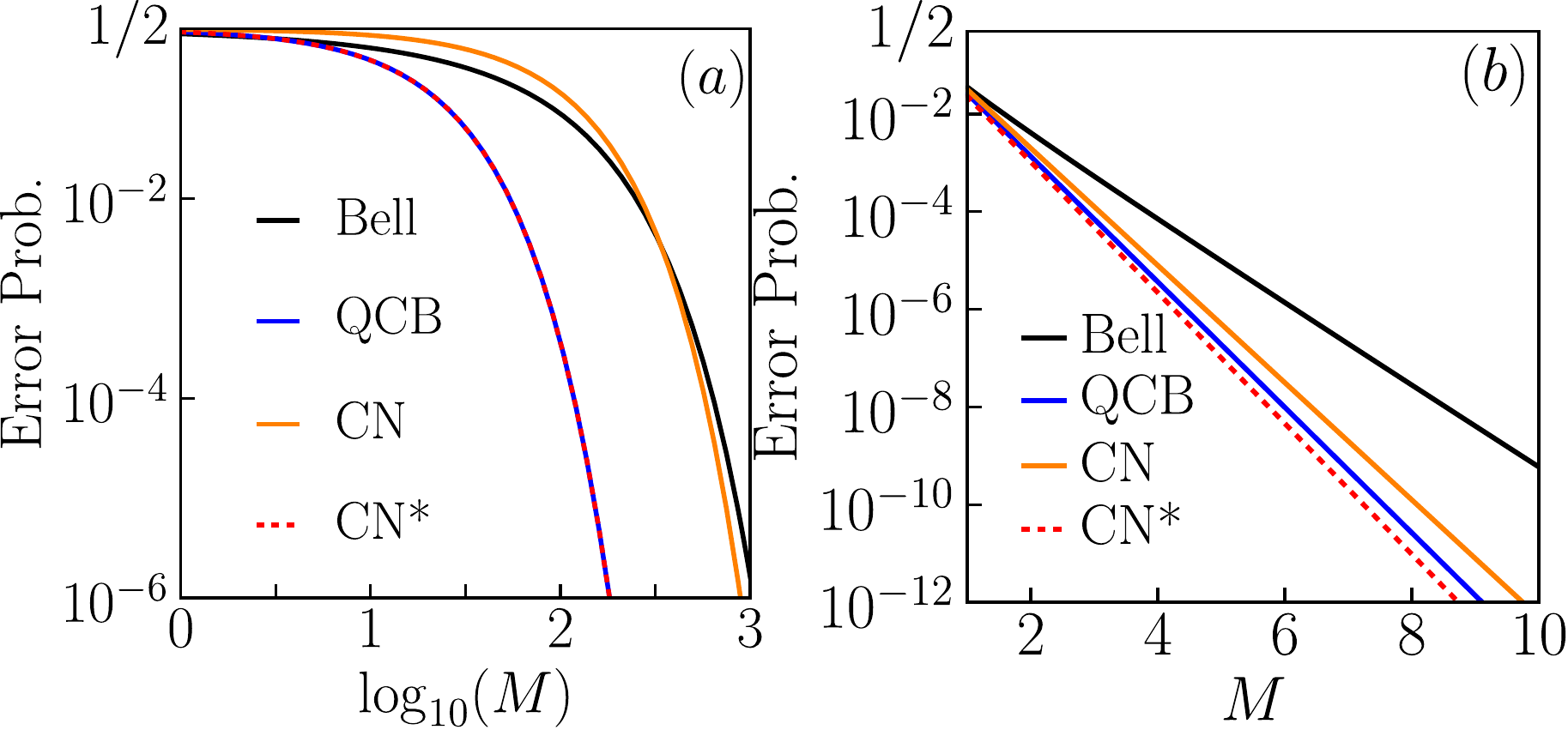}
\caption{{\bf Error probability versus number of modes $M$ for binary quantum reading}. \QZnew{Background and target reflectivities} are respectively $r_B=1$ and $r_T=0.4$. Comparisons are done for \QZnew{a number of photons per mode} $N_S=0.1$ in panel~(a) and $N_S=10$ in panel~(b). We plot the performance of the original Bell receiver~\cite{pirandola2011quantum} (solid black line), the asymptotically tight quantum Chernoff bound (QCB, solid blue line), the generalized \QZnew{conditional nulling receiver with performance $\zeta_1^{\rm QR}/2$ (CN, solid orange line), and the generalized conditional nulling receiver with improved performance $\zeta_{1*}^{\rm QR}/2$ (CN*, red dashed line).}
\label{fig:QCB_QR}
}
\end{figure}

\QZ{
The novel CN receiver design also provides a new insight into the original quantum reading model, related to the binary discrimination between the two lossy channels $\calL_{r_T}$ and $\calL_{r_B}$. With no loss of generality, let us assume $r_B>r_T$. When the two-mode squeezed vacuum state is used at the input, the corresponding outputs for the two channels are  $\Xi^{(T)}$ and $\Xi^{(B)}$. Therefore, the t-POVM and b-POVM can be directly used to perform their discrimination, leading to the error probability $\zeta_1^{\rm QR}/2$ for equal prior probabilities, \SP{where $\zeta_1^{\rm QR}$ is given in Eq.~(\ref{QRM1})} (see orange line in Fig.~\ref{fig:QCB_QR}). In the ideal case of $r_B=1$, the further improved detection, given by the application of the CN receiver to both signals and idlers, leads to the error probability $\zeta_{1*}^{\rm QR}/2$, \SP{where $\zeta_{1*}^{\rm QR}$ is defined in Eq.~(\ref{QRM2})} (see red dotted line in Fig.~\ref{fig:QCB_QR}). We see that the improved performance $\zeta_{1*}^{\rm QR}/2$ saturates the quantum Chernoff bound~\cite{Audenaert2007,pirandola2008computable}, while the general applicable performance $\zeta_1^{\rm QR}/2$ is able to beat the best known Bell-measurement receiver designed in Ref.~\cite{pirandola2011quantum}, when $M$ is sufficiently large (Fig.~\ref{fig:QCB_QR}a) or $N_S$ is large (Fig.~\ref{fig:QCB_QR}b).
}

\subsection{Quantum-enhanced frequency scanner in noisy conditions}
\QZ{
The previous result on quantum-enhanced target finding can be mapped into the model of quantum-enhanced frequency scanner, now in the presence of bright environmental noise. Here we assume a target at some fixed linear distance which only reflects radiation at a narrow bandwidth $\delta\nu$ around some carrier frequency. The target is assumed to be still (or slowly moving) and it completely diffracts the other frequencies. This limited reflection could also be the effect of meta-materials employed in a cloak. The previous $m$ sectors now become $m$ different non-overlapping frequency windows with bandwidth $\delta\nu$, each of them probed by pulses with the same bandwidth. 
}

\QZ{
One choice is to use a single $\delta\nu$-pulse per window containing $M \simeq \delta\nu^{-1}$ effective frequencies, each with $N_S$ mean number of photons. Alternatively, we may use $M$ $\delta\nu$-pulses per window which are irradiated as a train of independent temporal modes, each with $N_S$ mean photons. In our basic model, reflection occurs in only one of these frequency windows, while background thermal noise is detected for all the other windows. The previous results (see Fig.~\ref{fig:PCN_QI}) automatically imply that the use of an entangled source outperforms any classical strategies in the regime of few photon numbers per mode.
}

\subsection{About target ranging} 
\label{me:ranging}
\QZ{
In quantum target finding, if we consider time bins instead of spatial bins, we can map
the problem of direction finding into that of ranging. However, at fixed direction but unknown distance, there is a crucial problem which makes the entangled strategy problematic. We must in fact ensure that the returning signal (if any) is combined with the corresponding idler. Since we do not know, a priori, the round-trip time from the target, we cannot synchronize signal and idler in a joint detection. A potential way around this issue is to generate a train of $m$ signal-idler pulses with well-separated carrier frequencies (e.g., with a bandwidth larger than the maximum Doppler shift from the target). Signal-idler pulses with different carrier frequencies are then jointly detected at the different $m$ time bins. In principle this procedure can make the quantum measurement work but it opens another issue. The best classical strategy does not need to employ this time slicing approach. In fact, one could just send a single coherent pulse and wait for its potential return. From an energetic point of view, the classical source would only irradiate $MN_S$ photons (assuming $M$ modes per pulse) while the quantum case needs to irradiate $mMN_S$ photons on the target. Taking into account of this difference, we cannot directly apply our previous findings and derive a conclusive result for target ranging.
}

\subsection{Optimality of pure states}
\label{me:pure}

\QZ{Here we state two lemmas to summarize the results (See Sec.~\ref{supp2} for their proofs).}

\begin{lemma}
\label{lemma:pure_state}
Consider the discrimination of $N$ channels $\{\calE_n\}$ with prior probabilities $\{p_n\}$. Inputting pure states minimizes the mean error probability.
\end{lemma}

\smallskip
Note that if there is a constraint on the Hilbert space (e.g., an energy constraint for an infinite-dimensional space), then the previous lemma might not hold. However, this result may still hold in the presence of convexity properties, as in the proof of the following lemma.

\begin{lemma}
\label{lemma:QR_classical}
Consider position-based quantum reading, with a constraint of $MN_S$ mean photon numbers per cell. Any statistical mixture of GUS coherent states can be reduced to $\otimes_{k=1}^m \ket{\alpha}_{S_k}$ with amplitude $\alpha=\sqrt{MN_S}$. The minimum error probability is 
\be
P_{H}^{\rm CR}(r_B,r_T,M,N_S)=P_H(m,\zeta^{\rm CR}),
\ee
where 
$ 
\zeta^{\rm CR}=|\braket{\sqrt{r_B}\alpha|\sqrt{r_T}\alpha}|^2=e^{-M N_S(\sqrt{r_B}-\sqrt{r_T})^2}
$
and the function $P_H$ is given in Eq.~(\ref{P_H_PPM}) of the main text.
\end{lemma}

\subsection{Generalized CN Receiver (proof of theorem~\ref{lemma:CN})}
\label{me:proof_theorem1}
Let us describe the measurement process starting from $n=1$, i.e., by checking the hypothesis $h_1$ that the target state $\sigma^{(T)}$ is in subsystem $S_1$. If $h_1$ is true, then the receiver will not make any error, due to $\tr\small(\Pi_t^{(T)}\sigma^{(T)}\small)=1$ on the first subsystem $S_1$ and $\tr\small(\Pi_t^{(B)}\sigma^{(B)}\small)=1$ on all the other subsystems $\{S_k\}_{k=2}^m$. There is an error only if the true hypothesis is one of $\{h_k\}_{k=2}^m$. In this case, $S_1$ would be in the background state $\sigma^{(B)}$ and the $t$-POVM $\{\Pi_t^{(T)},\Pi_t^{(B)}\}$ would return the incorrect outcome `T' with probability $\zeta_1$ and correct outcome `B' with probability $1-\zeta_1$. 

Suppose that we get `T' (with type-I false-positive probability $\zeta_1$) while the correct hypothesis is $h_{\tilde{k}}$ for some $\tilde{k} > 1$. In measuring the remaining subsystems $\{S_k\}_{k=2}^m$ in the $b$-POVM $\{\Pi_b^{(T)},\Pi_b^{(B)} \}$, the outcomes will be certainly equal to `B' for all systems with $k \neq \tilde{k}$ since they will all be in a background state $\sigma^{(B)}$. However, the application of $b$-POVM over the target state $\sigma^{(T)}$ of subsystem $S_{\tilde{k}}$ could give the wrong outcome `B' with type-II (false-negative) probability $\zeta_2$. If this happens the receiver would select the false hypothesis $h_1$. In this case, the overall (conditional) probability of error is given by the product of the two incorrect outcomes $\zeta_1\zeta_2$ times the probability that $h_1$ is false, i.e., $(m-1)m^{-1}$. Therefore, we get $P_{\not h_1}^{T}=(m-1)m^{-1}\zeta_1\zeta_2$.

Suppose that, from the first measurement, we instead get the correct outcome `B' (with probability $1-\zeta_1$). Then, the receiver would correctly discard the false hypothesis $h_1$ and would check the next one $h_2$. Denote by $P_{m-1}$ the total error probability of the receiver in distinguishing the remaining $m-1$ hypotheses. Then, the overall (conditional) probability of error is given by the product of $P_{m-1}$, and the joint probability of outcome `B' for $h_1$ being false. Therefore, we have  $P_{\not h_1}^{B}=(m-1)m^{-1}(1-\zeta_1) P_{m-1}$. If $m=2$, then in this case there is only one hypothesis left, and we have the initial condition $P_1=0$.

Overall, the error probability of the receiver $P_m \equiv P_m^{\rm CN}(\zeta_1,\zeta_2)$ will be equal to the sum of $P_{\not h_1}^{T}$ and $P_{\not h_1}^{B}$, so that we have the recursive formula
\be
P_m=\frac{m-1}{m}\left[(1-\zeta_1)P_{m-1}+\zeta_1\zeta_2 \right].
\ee
The initial conditions of the recursion is that
$P_1=0$ and $P_2=\zeta_1\zeta_2/2$. To solve the recursion, let us set $P_m=-g_m/m$ so that we have 
$(1-\zeta_1)g_{m-1}-g_m=(m-1)\zeta_1\zeta_2$ with initial conditions $g_1=0$ and $g_2=-\zeta_1\zeta_2$. We find the solution 
\begin{align}
& g_m=-\zeta_1\zeta_2 \sum_{n=1}^{m-2}(m-n)(1-\zeta_1)^{n-1}=\nonumber \\
& -\zeta_1\zeta_2(m\zeta_1+(1-\zeta_1)^m-1)/\zeta_1^2,
\end{align}
which leads to
\be
P_m=\frac{1}{m}\frac{\zeta_2}{\zeta_1}\left[m\zeta_1+(1-\zeta_1)^m-1\right],
\ee
completing the proof. Note that, when the receiver outcomes are all `B', this automatically means that the true hypothesis is the last one $h_m$, which is compatible with the initial condition $P_1=0$.

\subsection{General bounds}
\label{me:bounds}
Here we present various general bounds that apply to $m$-ary state discrimination (in the setting of symmetric hypothesis testing)~\cite{Bagan,Qiu,Barnum,Ogawa}. These bounds apply to the mean error probability and can be computed from the quantum fidelity (which has a closed formula for arbitrary multimode Gaussian states~\cite{banchi2015quantum}). In particular, for any ensemble of $m$
mixed states $\{p_{k},\rho_{k}\}_{k=1}^{m}$, where $p_k$'s are the prior probabilities and $\rho_k$'s are the states, we may write the following upper
bound~\cite{Barnum} on the minimum error probability or Helstrom limit $P_H$
\begin{equation}
P_H\leq P_{H,UB}\equiv 2 \sum_{k^\prime>k}\sqrt{p_{k^\prime}p_{k}}F(\rho_{k^\prime},\rho_{k}),
\label{Barnum}
\end{equation}
where $F$ is the Bures' fidelity 
\be
F(\rho,\sigma):=\Vert\sqrt{\rho}\sqrt{\sigma}\Vert_{1} =\tr \sqrt{\sqrt{\rho}\sigma\sqrt{\rho}}.
\ee
The result of Eq.~(\ref{Barnum}) is a bound on the performance of a `pretty
good' measurement~\cite{PGM1,PGM2,PGM3} and is tight up to constant factors in the exponent. A fidelity-based lower bound is instead given by~\cite{montanaro2008lower},
\begin{equation}
P_H\geq P_{H,LB}\equiv \sum_{k^\prime>k}p_{k^\prime}p_{k}F^{2}(\rho_{k^\prime},\rho_{k}).
\label{montanaro2008lower}%
\end{equation}

Assume equi-probable hypotheses, so that $p_{k}=m^{-1}$ for any $k$, and the symmetry $F(\rho_k,\rho_{k^\prime})=F, \forall k\neq k^\prime$. We then have the simplified bounds
\begin{align}
 P_{H,UB}\equiv (m-1)F,
\label{UB2}\\
 P_{H,LB}\equiv \frac{m-1}{2m}F^2. 
\label{LB2}
\end{align}
These bounds appear in our main text with the following expressions for the fidelity
\begin{equation}
F(\rho_n,\rho_{n^\prime\neq n})=F^2\big(\Xi^{(T)},\Xi^{(B)}\big),
\end{equation}
for the entangled case and 
\be
F(\rho_n,\rho_{n^\prime\neq n})=F^2\big(\sigma^{(T)},\sigma^{(B)}\big)
\ee 
for the classical case.

\subsection{Classical benchmarks}
\label{App:classical_benchmarks} 
Let us now introduce a general bound to the ultimate performances achievable by classical states in CPF, with direct application to the problems of position-based reading and target finding. Recall that the general problem of CPF consists of discriminating an ensemble of GUS bosonic channels $\{\calE_n\}$ with equal priors. These are expressed by
\be
\calE_n=\big(\otimes_{k\neq n} \Phi^{(B)}_{S_k}\big)\otimes \Phi^{(T)}_{S_n},
\label{channel_GUS2}
\ee 
where $\Phi^{(B/T)}_{S_k}$ is the background/target channel acting on subsystem $S_k$ (e.g., a cell or a sector). Each of these channels is generally meant to be a multi-mode channel.  

In the bosonic setting, single-mode phase insensitive Gaussian channels model various physical processes. This channel $\mathcal{G}_{\mu,E}$ can be parameterized by a transmissivity/gain parameter $\mu>0$ and a noise parameter $E>0$~\cite{holevo2007one,weedbrook2012gaussian}. In particular, $E$ accounts for the thermal photons at the output of the channel, when the input state is a vacuum or coherent state. 
Besides the single-mode phase-insensitive (covariant) bosonic Gaussian channels \SP{discussed above}, we can also include the contravariant conjugate thermal-amplifier channel, whose action on an input annihilation operator is described by
\be 
\hat{a}\to \sqrt{\mu} \hat{a}^\dagger+\sqrt{\mu+1}\hat{e},
\ee 
where $\mu>0$ and $\hat{e}$ is in a thermal state with mean photon number $(E-\mu)/(\mu+1)$. All these channels $\mathcal{G}_{\mu,E}$ map a coherent state $\ket{\alpha}$ to a displaced thermal state with amplitude $\sqrt{\mu}\alpha$ ($\sqrt{\mu}\alpha^\star$ for the conjugate thermal-amplifier channel) and covariance matrix $(2E+1)\bm I$.

Therefore, let us consider the problem of CPF where target and background channels are tensor products of a phase-insensitive bosonic Gaussian channel $\mathcal{G}_{\mu,E}$. Denote the transmissivity/gain and noise of the target channel as $\mu_T$ and $E_T$, while those of the background channel as $\mu_B$ and $E_B$. For the entangled case, we assume that each subsystem is exactly probed by $M$ signal modes, each irradiating $N_S$ mean photons, for a total of $mMN_S$ mean photons. For the classical case, we can relax this structure and include the more general case of different energies irradiated by the $M$ modes over each subsystem $S_k$. More generally, for the classical case with no passive signature ($E_B=E_T$), we can also allow for arbitrary number of modes $M_{k}$ per subsystem $S_k$ so that $\Phi_{S_k}^{(l)}=\mathcal{G}_{\mu_{l},E_{l}}^{\otimes M_k}$. In other words, for classical CPF with no passive signature, the only surviving constraint is the $mMN_S$ mean photons globally irradiated. More precisely, we can state the following result \QZ{(See Sec.~\ref{supp2} for proof)}.

\begin{lemma}
\label{lemma:joint_lower_bound}
Consider the problem of CPF where target and background channels are tensor products of a single-mode phase-insensitive bosonic Gaussian channel with parameters $\mu_T, E_T$ (for target) and $\mu_B, E_B$ (for background). Assume a global energetic constraint of $mMN_S$ mean photons with $M$ modes irradiated over each of the $m$ subsystems $S_k$. The optimal classical state (with positive P-representation) minimizing the lower bound $P_{H,LB}$ of Eq.~(\ref{montanaro2008lower}) is any tensor product of coherent states
\be
\ket{\bm \alpha}=\otimes_{k=1}^{m} \big(\otimes_{k^\prime=1}^M\ket{e^{i\theta_k^{(k^\prime)}}\sqrt{N_{S_k}^{(k^\prime)}}}\big)_{S_k},
\ee
where the phases $\theta_k^{(k^\prime)}$ are arbitrary and $\sum_{k^\prime=1}^M N_{S_k}^{(k^\prime)}=MN_S$ for any $k$, so that each subsystem is irradiated by the same mean number of photons. The corresponding minimum lower bound is given by
\begin{align}
P_{H,LB} \equiv \frac{m-1}{2m} & c_{E_B,E_T}^{2M} \times \nonumber \\ & \exp\left[-\frac{2MN_S(\sqrt{\mu_B}-\sqrt{\mu_T})^2}{1+E_B+E_T} \right], 
\label{LB_Gaussi_app}
\end{align}
with $c_{E_B,E_T}=[1+\big(\sqrt{E_B(1+E_T)}-\sqrt{E_T(1+E_B)}\big)^2]^{-1}$. In particular, for no passive signature ($E_T=E_B \equiv E$), we have the simplification 
\be
P_{H,LB}\equiv \frac{m-1}{2m} \exp\left[-\frac{2MN_S(\sqrt{\mu_B}-\sqrt{\mu_T})^2}{1+2 E} \right],
\label{LB_Gaussi_app2}
\ee
and bound holds under the general energetic constraint of $mMN_S$ mean photons, with no restriction on the number of modes irradiated per subsystem. In this case, an optimal state is the tensor-product $\otimes_{k=1}^{m} \ket{\sqrt{MN_S}}_{S_k}$.
\end{lemma}

\subsection{Position-based quantum reading with thermal noise}
\label{noise_reading}
Let us now generalize the study of position-based quantum reading to the case where thermal noise is present in the environment. This means that the environmental input of each cell $S_k$ is not the vacuum but a thermal state with $N_B$ mean photons. Each cell has reflectivity $r_B$ or $r_T$ in such a way that the block of $m$ cells has GUS. The block is probed by bosonic modes for a total of $mMN_S$ mean photons irradiated. In the classical case, we compute a lower bound to the performance of all possible classical states (globally irradiating $mMN_S$ mean photons over the $m$ block of cells), while for the quantum case, we consider a tensor-product of two-mode squeezed vacuum states, so that $M$ signal modes probe each cell, with each mode irradiating $N_S$ mean photons. 

As before, this problem is mapped into the discrimination of an ensemble of GUS bosonic channels $\{\calE_n\}$ with equal priors, which are expressed by
\be
\calE_n=\big(\otimes_{k\neq n} \Phi^{(B)}_{S_k}\big)\otimes \Phi^{(T)}_{S_n},
\label{channel_GUS3}
\ee 
with $\Phi^{(B/T)}_{S_k}$ acting on cell $S_k$. For $M$-mode probing of the cell, we have the target channel $\Phi^{(T)}={\big(\mathcal{L}_{r_T}^{N_B}\big)}^{\otimes M}$ and the background channel 
$\Phi^{(B)}={\big(\mathcal{L}_{r_B}^{N_B}\big)}^{\otimes M}$, where $\mathcal{L}_{r}^{N_B}$ is a single-mode thermal-loss channel with reflectivity $r$ and thermal noise $N_B$.

In general, the protocol of position-based quantum reading can be formulated with two generic thermal-loss channels as discussed above. In such a case, the classical benchmark can be easily derived from Eq.~(\ref{LB_Gaussi_app}). Then, we may introduce a finer classification of the protocol in two types: one with active and the other with passive signature. In the first type of protocol, the parameters of the channels are such that the noise variance at the output of the two channels is different assuming the vacuum state at the input. In other words, their statistical discrimination is possible without sending a probing signal. In the second type, the parameters are such that there are no different levels of noise at the output. Here we analyze this second type, so that the channels have reflectivity $r_l$ and mean number of thermal photons $N_B/(1-r_l)$ for $l=B,T$. The corresponding classical benchmark can be computed from Eq.~(\ref{LB_Gaussi_app2}) and takes the form  
\begin{align} 
&
P_{H,LB}^{\rm CR,N}(r_B,r_T,M,N_S)=
\nonumber
\\
&
\frac{m-1}{2m}\exp\left[\frac{-2M(\sqrt{r_B}-\sqrt{r_T})^2N_S}{2N_B+1}\right].
\label{LB_CR_NB}
\end{align}

Similarly, for the quantum case, we can easily repeat the calculations to find the corresponding noisy expression $P_{H,UB}^{\rm QR,N}$ of the upper bound $P_{H,UB}^{\rm QR}$. For $N_S\ll1$ and $M\gg1$ at fixed $MN_S$, we may generalize Eq.~(\ref{P_H_QR_E_UB_asym}) of our main text into the following form
\begin{align}
&P_{H,UB}^{\rm QR,N}(r_B,r_T,M,N_S)\simeq 
\nonumber
\\
&(m-1)\exp\left[\frac{-2MN_S(1+N_B-\sqrt{H}-\sqrt{r_Br_T})}{1+N_B}\right],
\label{P_H_QR_E_UB_asym_NB}
\end{align} 
where $H=(1+N_B-r_B)(1+N_B-r_T)$.

Denote the error exponent in Eq.~(\ref{LB_CR_NB}) as $\epsilon_{CR}$ and the error exponent in Eq.~(\ref{P_H_QR_E_UB_asym_NB}) as $\epsilon_{QR}$.
We find that the quantum case is always better than the classical case, i.e. $\epsilon_{QR}>\epsilon_{CR}$. For $r_T$ and $r_B$ close to $1$, we have $\epsilon_{QR}/\epsilon_{CR}\simeq 1+1/2N_B$. In this regime, we see that the advantage becomes huge when $N_B\ll1$, which agrees with our observation in Eqs.~(\ref{P_H_QR_Classical_asym}) and~(\ref{P_H_QR_E_UB_asym}). However, when $N_B\gg1$, the advantage decays, in agreement with the observation related to Eqs.~(\ref{P_H_CI_LB}) and~(\ref{P_H_QI_E_UB_asym}). Note that this conclusion is based on a quantum lower bound and a classical upper bound, and we expect them to be not tight when noise $N_B$ is large.

\bigskip
\noindent{\bf \normalsize Acknowledgments}\\
Q.Z. thanks discussions with Zheshen Zhang on possible experiments. S.P. thanks discussions with Gae Spedalieri. Q.Z. acknowledges funding from Army Research Office under Grant Number W911NF-19-1-0418 and University of Arizona. S.P. acknowledges funding from the European Union's Horizon 2020 Research and Innovation Action under grant agreement No. 862644 (Quantum readout techniques and technologies, QUARTET). 

\bigskip
\noindent{\bf \normalsize Author Contributions}\\
SP formulated the problem of `channel-position finding' and the general methodology for its investigation. SP proposed the applications for quantum reading and spectroscopy, while target finding and ranging was jointly proposed by the two authors. QZ conceived the analysis via the geometric uniform symmetry, designed the generalized conditional nulling receiver (with refinements introduced by SP) and developed the corresponding analytical derivations (with contributions from SP). QZ performed the numerical evaluations and generated the plots. Both authors wrote the manuscript.

\section{Supplementary Note}
\subsection{Supplementary Note 1: Fidelity calculations for bosonic Gaussian states} 
\label{supp1}

Given two arbitrary multimode Gaussian states, we can compute their fidelity using the general analytical formula of Ref.~[R1]. When the states are one-mode or two-mode, this formula reduces to the results of Refs.~[R2,R3]. Here we use these tools for our calculations.

Given two Gaussian states $\rho_\ell$ ($\ell=1,2$), we denote their means and covariance matrices as $\bm u_\ell$ and  $\mathbf{V}_\ell$. We also define the difference $\delta \bm u=\bm u_2-\bm u_1$. 
For the one mode case, let $\Delta w=\text{Det}[\mathbf{V}_1+\mathbf{V}_2]$, $\delta w=\left(\text{Det}[\mathbf{V}_1]-1\right)\left(\text{Det}[\mathbf{V}_2]-1\right)$, we have the fidelity 
\begin{align}
&F^2\left[\rho_1,\rho_2\right]=
\frac{2\exp\left[-\frac{1}{2}\delta \bm u^T (\mathbf{V}_1+\mathbf{V}_2)^{-1} \delta \bm u\right]}{\sqrt{\Delta w+\delta w}-\sqrt{\delta w}}.
\label{fidelity_one_mode}
\end{align}
For the two-mode case, define $\mathbf{J}=i{\bm Z}_2\bigoplus i{\bm Z}_2$, where ${\bm Z}_2$ is the two-by-two Pauli matrix. Then set $\Delta={\rm Det}[(\mathbf{V}_1+\mathbf{V}_2)/2]$,  $\Gamma=2^4 {\rm Det}[\mathbf{J}\mathbf{V}_1\mathbf{J}\mathbf{V}_2/4-\mathbf{I}_4/4]$ and $\Lambda=2^4 {\rm Det}[\mathbf{V}_1/2+i \mathbf{J}/2] {\rm Det}[\mathbf{V}_2/2+i \mathbf{J}/2]$. The fidelity is
\begin{align}
&F^2\left[\rho_1,\rho_2\right]
=\frac{\exp\left[-\frac{1}{2}\delta \bm u^T (\mathbf{V}_1+\mathbf{V}_2)^{-1} \delta\bm u \right]}{\sqrt{\Gamma}+\sqrt{\Lambda}-\sqrt{(\sqrt{\Gamma}+\sqrt{\Lambda})^2-\Delta}}.
\label{fidelity_two_mode}
\end{align}


The two-mode squeezed vacuum state $\phi_{\rm ME}$ has covariance matrix
\begin{align}
&
{\mathbf{V}}_{\rm ME}=
\left(
\begin{array}{cccc}
(2N_S+1) {\mathbf I}_2&2C_p{\mathbf Z}_2\\
2C_p{\mathbf Z}_2&(2N_S+1){\mathbf I}_2
\end{array} 
\right),
\label{V_TMSS}
&
\end{align}
where $C_p=\sqrt{N_S\left(N_S+1\right)}$ and ${\mathbf Z}_2$ is the Pauli Z matrix.

For for position-based quantum reading, after pure-loss affects the signal mode, we have the joint return-idler state $(\calL_r\otimes I) \phi_{\rm ME}$, with zero mean and covariance matrix
\begin{align}
&
{\mathbf{V}}_{QR}(r) =
\left(
\begin{array}{cccc}
(2rN_S+1) {\mathbf I}_2&2\sqrt{r}C_p{\mathbf Z}_2\\
2\sqrt{r}C_p{\mathbf Z}_2&(2N_S+1){\mathbf I}_2
\end{array} 
\right).
\label{cov_QR}
\end{align}
The two states for quantum reading are zero mean and covariance matrix ${\mathbf{V}}_1={\mathbf{V}}_{QR}(r_T)$ and ${\mathbf{V}}_2={\mathbf{V}}_{QR}(r_B)$. The fidelity can be evaluated using Eq.~(\ref{fidelity_two_mode}).
Consider also the noisy version of the protocol where the cells are subject to environmental thermal noise, with $N_B>0$ mean photons per mode. Then, the covariance matrix in Eq.~(\ref{cov_QR}) changes to
\begin{align}
&
{\mathbf{V}}_{QR}(r) =
\left(
\begin{array}{cccc}
(2rN_S+2N_B+1) {\mathbf I}_2&2\sqrt{r}C_p{\mathbf Z}_2\\
2\sqrt{r}C_p{\mathbf Z}_2&(2N_S+1){\mathbf I}_2
\end{array} 
\right).
\label{cov_QR_NB}
\end{align}

For quantum target finding, after the target channel, we have $(\calL_\eta^{N_B/(1-\eta)}\otimes \calI) \phi_{\rm ME}$ with zero mean and covariance matrix
\begin{align}
&
{\mathbf{V}}_1=
\left(
\begin{array}{cccc}
(2N_B+1) {\mathbf I}_2&2\sqrt{\eta}C_p{\mathbf Z}_2\\
2\sqrt{\eta}C_p{\mathbf Z}_2&(2N_S+1){\mathbf I}_2
\end{array} 
\right).
\label{noisy_cov}
&
\end{align}
At the output of the background channel, we have a state $(\calL_0^{N_B}\otimes \calI) \phi_{\rm ME}$ with zero mean and covariance matrix
\begin{align}
&
{\mathbf{V}}_2=
\left(
\begin{array}{cccc}
(2N_B+1) {\mathbf I}_2&0\\
0&(2N_S+1){\mathbf I}_2
\end{array} 
\right).
&
\end{align}
The fidelity can be evaluated using Eq.~(\ref{fidelity_two_mode}).

\subsection{Supplementary Note 2: Proofs of lemmas}
\label{supp2}

\subsubsection{Proof of Lemma~\ref{lemma:pure_state}}

Suppose that $\{\Pi_n\}$ is the positive-operator valued measure (POVM) for discriminating the channels $\{\calE_n\}$ (with prior probability $\{p_n\}$), then the error probability, given an input state $\rho \in \calH$ is
\be
P_E(\rho, \{\calE_n\},\{\Pi_n\})=1-\sum_n p_n \tr\left[\calE_n(\rho) \Pi_n\right].
\ee   
Optimizing over the POVMs, the minimum error probability for a given input state $\rho$ is 
\ba
P_H(\rho, \{\calE_n\})&=&\min_{\{\Pi_n\}} P_E(\rho, \{\calE_n\},\{\Pi_n\})
\nonumber
\\
&=&1-\max_{\{\Pi_n\}} \sum_n p_n \tr\left[\calE_n(\rho) \Pi_n\right].
\ea  
Perform the spectral decomposition $\rho=\sum_\ell \lambda_\ell \psi_\ell$, where $\psi_\ell$'s are pure states and $\lambda_\ell$ are probabilities, $\sum_\ell \lambda_\ell=1$. Then, we derive
\begin{align} 
&P_H(\rho, \{\calE_n\})=1-\max_{\{\Pi_n\}}\sum_{\ell}\lambda_\ell \sum_n p_n \tr\left[\calE_n(\psi_\ell) \Pi_n\right] \nonumber
\\
&\ge \sum_{\ell}\lambda_\ell \left[1- \max_{\{\Pi_n^\prime\}} \sum_n p_n \tr\left[\calE_n(\psi_\ell) \Pi_n^\prime\right]\right] \nonumber
\\
&=\sum_{\ell}\lambda_\ell P_H(\psi_\ell,\{\calE_n\}) \nonumber
\\
&\ge  \min_{\psi} P_H(\psi, \{\calE_n\}),
\end{align}
which completes the proof. Note that, in the second step, we interchanged the order of maximization and summation, which leads to an inequality. Also note that the optimum $\{\Pi_n^\prime\}$ in the second line can be $\ell$-dependent.

\subsubsection{Proof of Lemma~\ref{lemma:QR_classical}}

Let us start by considering the GUS tensor-product $\ket{\bm \beta}=\otimes_{k=1}^m \ket{\bm \alpha}_{S_k}$, where each multi-mode coherent state for subsystem $\ket{\bm \alpha}=\otimes_{k=1}^M \ket{\alpha_k}$ has arbitrary $M$ modes and a total of $n_S=\bm \alpha^\dagger \bm \alpha=|\bm \alpha|^2$ mean photons. 
Each subsystem is subject to a pure-loss channel $\calL_{r_\ell}^{\otimes M}$ with reflectivity $r_\ell$, giving the output $\calL_{r_\ell}^{\otimes M}(\ket{\bm \alpha})=\ket{\sqrt{r_\ell}\bm \alpha}$ (where $\ell=B,T$ depends on the cell, i.e., if it is a background or a target cell). Given the input source $\ket{\bm \beta}$, the total GUS state at the output of the global channel $\calE_n$ is therefore given by
\be
\ket{\psi_n}=\big(\otimes_{k\neq n}\ket{\sqrt{r_B}\bm \alpha}_{S_k}\big)\otimes \ket{\sqrt{r_T}\bm \alpha}_{S_n}.
\label{app_output}
\ee 

The pulse-position modulation discrimination of this GUS ensemble of pure states is affected by a minimum error probability $P_H(\ket{\bm \beta}\bra{\bm \beta},\{\calE_n\})$, which is given by the Helstrom limit $P_{H}(m,\zeta^{\rm CR})$ in Eq.~(\ref{P_H_PPM}) of the main text, with the following overlap (or fidelity)
\be \zeta_{n_S}=|\braket{\sqrt{r_B}\bm \alpha|\sqrt{r_T}\bm \alpha}|^2=e^{-(\sqrt{r_B}-\sqrt{r_T})^2 n_S}.
\ee 
Here it becomes clear that we can simply consider $M=1$, without loss of generality, since $M$ does not appear in the Helstrom limit above. In fact, a general $M$-mode coherent state in each subsystem in Eq.~(\ref{app_output}) can be unitarily mapped into a single mode coherent state with the same energy, leaving all other $M-1$ modes in vacua.

Consider now a GUS classical state that can be written as the following mixture of coherent states
\be
\rho_C=\int{d\nu}  \left[\otimes_{k=1}^m \ket{\sqrt{n_S}e^{i\theta}}\bra{\sqrt{n_S}e^{i\theta}}\right],
\ee
where $d\nu$ represents the Lebesgue integral with an arbitrary probability measure $\nu$ over $n_S$ and $\theta$, i.e., 
\be
\int d \nu f(n_S,\theta)=\int dn_S d\theta~\nu(n_S,\theta)  f(n_S,\theta)
\ee
with the probability density function $\nu(n_S,\theta)$.
The mean number of photons per subsystem of the state is given by
\be
\bar{n}_S  =  \int d\nu^\prime n_S  \le MN_S,
\label{constraint_mp}
\ee
where $\int{d\nu}$ has been reduced to $\int{d\nu^\prime}$ with the marginal probability measure $\nu^\prime$ restricted to the variable $n_S$.

Now, following the proof of Lemma~\ref{lemma:pure_state}, we write
\begin{align}
& P_H \left(\rho_C, \{\calE_n\} \right) 
\nonumber
\\
& \ge \int{d\nu} P_H(\otimes_{k=1}^m \ket{\sqrt{n_S}e^{i\theta}}\bra{\sqrt{n_S}e^{i\theta}},\{\calE_n\}) \nonumber \\
&= \int{d\nu^\prime} P_H(m,\zeta_{n_S}).
\label{Aeq}
\end{align}
Moreover, because $\zeta_{n_S}$ does not depend on $\theta$, the integral above has been restricted to a marginal probability measure $\nu^\prime$ over $n_S$. Note that $P_H(m,\zeta_{n_S})$ is a non-decreasing function of $\zeta_{n_S}$ and also convex in $\zeta_{n_S}$, which can be shown by calculating the first and second order derivatives. Then, $\zeta_{n_S}$ is a convex function in $n_S$. For a function $f$ convex in $x$ and another function $g$ convex and non-decreasing in $f$, we have that the composition $g[f(x)]$ is convex in $x$. This means that $P_H(m,\zeta_{n_S})$ is convex in $n_S$ and we can write
\ba
& \int{d\nu^\prime} P_H(m,\zeta_{n_S}) \ge P_H(m,\zeta_{\int{d\nu^\prime} n_S})
\nonumber
\\
&\ge P_H(m,\zeta_{MN_S})=P_H(m,\zeta^{\rm CR}),\label{toogeneral}
\ea
where, in the last inequality, we used the constraint in Eq.~(\ref{constraint_mp}) and the fact that $\zeta_{n_S}$ is decreasing in $n_S$. From Eq.~(\ref{toogeneral}) it is clear that the minimum error probability is achieved by the state $\otimes_{k=1}^m \ket{\sqrt{n_S}}_{S_k}$ in the original lower bound.

\subsubsection{Proof of Lemma~\ref{lemma:joint_lower_bound}} 

For the convenience of analysis, we will parameterize a coherent state $\ket{\alpha}$ with the phase and amplitude squared, i.e.,  $\ket{x,\theta}\equiv \ket{\sqrt{x}e^{i\theta}}$, where $x \ge 0$ and $0 \le \theta \le 2\pi$. In this notation, a multi-mode coherent state over the entire system takes the form $\ket{\bm x, \bm \theta}=\otimes_{k=1}^m \big(\ket{\bm x_k,\bm \theta_k}_{S_k}\big)$, where each subsystem state $\ket{\bm x_k,\bm \theta_k}_{S_k}=\otimes_{k^\prime=1}^M \ket{x_k^{(k^\prime)},\theta_k^{(k^\prime)}}$ is again a tensor product of multiple modes with generally-different amplitudes. Here $\bm x_k$ are positive and real vectors $\bm x_k=(x_k^{(1)},\cdots, x_k^{(M)})\equiv \{x_k^{(k^\prime)}\}_{k^\prime=1}^M$ and $\bm x$ is a simple concatenation of them, i.e., $\bm x=(\bm x_1,\cdots, \bm x_m)$.

In this notation, the general classical state
as the input can be written as a Lebesgue integral
\be
\rho=\int{dP}  \ket{\bm x, \bm \theta}\bra{\bm x, \bm \theta},
\ee 
where the probability measure $P$ over $\bm x,\bm \theta$ can be arbitrary. Let us define 
\be
\|\bm x\|_1\equiv\sum_{k,k^\prime} |x_k^{(k^\prime)}|=\sum_{k,k^\prime} x_k^{(k^\prime)},
\ee
which is the standard one-norm and equals the total mean photon number of the state $\ket{\bm x, \bm \theta}$. Then, the total energy constraint leads to the inequality
\be
\int{dP^\prime} \|\bm x\|_1 \le mMN_S,
\label{energy_constraint_one_norm}
\ee 
where the integral has been simplified to a marginal probability measure $P^\prime$ restricted to the non-negative variables $\bm x$. 

The total conditional state at the output of the channel $\calE_n$ is also a mixture, with expression
\begin{align}
\rho_n^{\rm C}=\calE_n  (\rho)=\int{dP} \rho^{\rm C}_{\bm x, \bm \theta,n},
\end{align}
where each conditional state is given by
\be
\rho^{\rm C}_{\bm x, \bm \theta,n}=
\big(\otimes_{k\neq n}(\rho^{\rm C}_B)_{S_k} \big)\otimes (\rho^{\rm C}_T)_{S_n}.
\ee
The target state $(\rho^{\rm C}_T)_{S_n}$ is a product of $M$ displaced thermal states, each with amplitude $\sqrt{\mu_Tx_n^{(k^\prime)}}e^{i\theta_n^{(k^\prime)}}$ and covariance matrix $(2E_T+1)\bm I$; the background state $(\rho^{\rm C}_B)_{S_k}$ is a product of other $M$ displaced thermal states, each with amplitude $\sqrt{\mu_B x_k^{(k^\prime)}}e^{i\theta_k^{(k^\prime)}}$ and covariance matrix $(2E_B+1)\bm I$.

From Eq.~(\ref{montanaro2008lower}), we can write the following lower bound to the mean error probability.
\begin{align}
&
P_{H,LB}^{\rm C}=\sum_{n^\prime>n}\frac{1}{m^2} F^2\left[\int{dP} 
\rho_{\bm x,\bm \theta,n}^{\rm C} ,\int{dP}
\rho_{\bm x,\bm \theta,n^\prime}^{\rm C} \right] \nonumber
\\
&
\ge 
\frac{K}{m^2}\sum_{n^\prime>n} \frac{1}{K}
\Big\{\int{dP}   F[
\rho_{\bm x,\bm \theta,n}^{\rm C},
\rho_{\bm x,\bm \theta,n^\prime}^{\rm C}] \Big\}^2 \nonumber
\\
&
\ge 
\frac{K}{m^2}
\Big\{ \sum_{n^\prime>n} \frac{1}{K}\int{dP}   F[\rho_{\bm x,\bm \theta,n}^{\rm C},
\rho_{\bm x,\bm \theta,n^\prime}^{\rm C}]\Big\}^2, \label{toreplace}
\end{align}
where use the joint concavity of the fidelity 
\be
F\left[\int{dp_x}  \rho_x ,\int{dp_x}  \sigma_x \right]\ge \int{dp_x}  F[\rho_x,\sigma_x],
\ee 
and Jensen's inequality for the square function with $K=(m-1)m/2$. 

Let us now address each fidelity term 
\begin{align}
&
F^{\rm C} \equiv F[\rho_{\bm x,\bm \theta, n}^{\rm C},\rho_{\bm x,\bm \theta,n^\prime\neq n}^{\rm C}] \nonumber
\\
&=
F\left[  (\rho^{\rm C}_T)_{S_n},(\rho^{\rm C}_B)_{S_n}\right] F\left[  (\rho^{\rm C}_B)_{S_{n^\prime}},(\rho^{\rm C}_T)_{S_{n^\prime}}\right].
\end{align}
Using Eq.~(\ref{fidelity_one_mode}), we can compute 
\be
F^{\rm C}=c_{E_B,E_T}^{M}\exp\left[-B (\|\bm x_n\|_1+\|\bm x_{n^\prime}\|_1) \right], 
\label{FC_general}
\ee
where the constant
$
B \equiv (\sqrt{\mu_B}-\sqrt{\mu_T})^2/(1+E_B+E_T)
$.
From the one-norm in the expression above, it becomes clear that the performance is exactly the same regardless how the energy is distributed among the $M$ modes impinging on a subsystem, as long as the mean total energy irradiated over the subsystem is fixed. 
By replacing the $F^{\rm C}$ in Eq.~(\ref{toreplace}), and noticing that $F^C$ does not depend on $\bm \theta$ we find the following lower bound
\begin{align}
P_{H,LB}^{C}
\ge 
\frac{c_{E_B,E_T}^{2M}K}{m^2} \Big\{  \int{dP^\prime}    g(\{\|\bm x_n\|_1\}_{n=1}^m) 
\Big\}^2,
\end{align}
where we define the function 
\be 
g(\{\|\bm x_n\|_1\}_{n=1}^m)\equiv  \frac{1}{K}\sum_{n^\prime>n}
\exp\left[ -B (\|\bm x_n\|_1+\|\bm x_{n^\prime}\|_1)\right].
\ee 
We notice that $e^{-cx}$ with $c\ge 0$
is strictly convex in the variable $x$. 
Thus, from convexity, we have
\ba
&&g(\{\|\bm x_n\|_1\}_{n=1}^m)
\nonumber
\\
&&
\ge 
\exp\left[-
\frac{1}{K}\sum_{n^\prime>n}
B (\|\bm x_n\|_1+\|\bm x_{n^\prime}\|_1)
\right] \nonumber
\\
&& 
=\exp\left[-
\frac{(m-1)}{K}\sum_{n=1}^m
B \|\bm x_n\|_1
\right] \nonumber
\\
&&= \exp\left[-2B\|\bm x\|_1/m\right],
\ea 
where we have used $K=m(m-1)/2$ and $\|\bm x\|_1 = \sum_{n=1}^m \|\bm x_n\|_1$ (from its definition). The equality holds if and only if $\|\bm x_n\|_1=\|\bm x\|_1/m$ for all $n$.

Thus overall we may write 
\begin{align}
&
P_{H,LB}^{C}
\ge
\frac{c_{E_B,E_T}^{2M} K}{m^2}
\Big\{  \int{dP^\prime}  \exp\left[-2B\|\bm x\|_1/m\right] \Big\}^2 \nonumber \\
&\ge
\frac{c_{E_B,E_T}^{2M} K}{m^2}
\Big\{\exp\left[ \int{dP^\prime} (-2B\|\bm x\|_1/m) \right]\Big\}^2 \nonumber
\\
&\ge
\frac{m-1}{2m} c_{E_B,E_T}^{2M}
\exp\left[-{2B M N_S }\right].
\end{align}
For the second inequality, we use the convexity of $e^{-c x}$ (with $c>0$) and Jensen's inequality to move expectation value to the exponent. The last inequality exploits the monotonic decreasing property of $e^{-c x}$ (with $c>0$) and the constraint in Eq.~(\ref{energy_constraint_one_norm}). This leads to the result in Eq.~(\ref{LB_Gaussi_app}) in the main paper. Due to Jensen's inequality and convexity, it is easy to check that the lower bound is reached by (and only by) an input coherent source $\ket{\bm x, \bm \theta}=\otimes_{k=1}^m \big(\ket{\bm x_k,\bm \theta_k}_{S_k}\big)$, such that on each subsystem the total mean photon number is equal, i.e., $\|\bm x_k\|=MN_S$.

In the passive case of $E_B=E_T \equiv E$, we have $c_{E_B,E_T}=1$, so that Eq.~(\ref{FC_general}) is replaced by
\be
F^{\rm C}=\exp\left[-B (\|\bm x_n\|_1+\|\bm x_{n^\prime}\|_1) \right].
\label{FC_passive}
\ee
We see that only the mean photon numbers of subsystems $S_n$ and $S_{n^\prime}$ appear in this expression, while the number of modes $M$ is no longer present. Following the same analysis from above we arrive at Eq.~(\ref{LB_Gaussi_app2}) in the main paper where the number of modes $M$ per subsystem can now be variable, as long as the total energetic constraint $mMN_S$ is fixed. In this case, the optimal state is a tensor product of coherent states with arbitrary number of modes per subsystem and arbitrary phases, and such to irradiate $MN_S$ mean photon number per subsystem. In particular, we may choose 
$\otimes_{k=1}^{m} \ket{\sqrt{MN_S}}_{S_k}$.

\end{document}